\documentclass[11pt]{article}
\pdfoutput=1
\usepackage[utf8]{inputenc}
\usepackage{multirow}
\usepackage{amsmath, amsfonts, amssymb}
\usepackage{array}
    \newcolumntype{C}[1]{>{\centering\arraybackslash$}p{#1}<{$}}
\usepackage{longtable}
\usepackage{comment}
\usepackage{graphicx}
\usepackage[table]{xcolor}
\usepackage{pdflscape}
\usepackage{psfrag}
\usepackage{amsthm}
\usepackage{enumerate}
\usepackage{arydshln}
\usepackage{pifont}
\usepackage{booktabs}

\usepackage{soul}
\usepackage{slashed}
\usepackage{subcaption}
\usepackage{mathrsfs}
\usepackage{ytableau}
 \usepackage{a4wide}
  \usepackage{tikz}
  \usepackage{tikz-cd}
  \usetikzlibrary{shapes.geometric}
  \usepackage{tcolorbox}
  \definecolor{dark-gray}{gray}{0.20}
  \definecolor{gray}{gray}{0.30}
  \definecolor{light-gray}{gray}{0.80}
  \definecolor{dark-red}{rgb}{0.7,0,0}
  \definecolor{dark-green}{rgb}{0.1,0.4,0}
  \definecolor{dark-blue}{rgb}{0.3,0.3,0.7}
  \definecolor{light-blue}{rgb}{0.8,0.8,1}
      \definecolor{swamp}{RGB}{240, 199, 197}

\usepackage{newunicodechar} 
\usepackage{setspace}

\usepackage{ifthen}

\usepackage{longtable}

\newcommand{\be}{\begin{equation}}
\newcommand{\ee}{\end{equation}}
\newcommand{\eq}[1]{(\ref{#1})}

\def\be{\begin{equation}}
\def\ee{\end{equation}}
\def\bea{\begin{eqnarray}}
\def\eea{\end{eqnarray}}

\newenvironment{eqaed}
    {\begin{equation}
    \begin{aligned}
    }
    { 
    \end{aligned}
    \end{equation}
    \ignorespacesafterend
    }

\newcommand{\Z}{\mathbb{Z}}
\AtBeginDocument{

}

\newcommand{\secref}[1]{Section~\ref{#1}}
\numberwithin{equation}{section}

\usepackage{enumitem}
\setlist[itemize,1]{leftmargin=3.8em}

\usepackage{jheppub}
\usepackage{hyperref}
\usepackage{varioref}  
\usepackage{cleveref}

\hypersetup{
	colorlinks=true,
	linkcolor=dark-blue,
	citecolor=dark-red,
	urlcolor=dark-green,
	linktoc=page
}

\theoremstyle{definition}

\theoremstyle{remark}

\crefname{appendix}{Appendix}{Appendices}

\usepackage{acronym}
\newacro{EFT}[EFT]{effective field theory}
\acrodefplural{EFT}{effective field theories}
\newacro{UV}[UV]{ultraviolet}
\newacro{KK}[KK]{Kaluza-Klein}
\newacro{BF}[BF]{Breitenlohner-Freedman}
\newacro{RFM}[RFM]{Riemann-flat manifold}
\acrodefplural{RFM}{Riemann-flat manifolds}
\newcommand{\M}{\mathcal{M}}
\newcommand{\C}{\mathcal{C}}
\newcommand{\D}[1][\gamma]{\mathbf{D}_{\mathfrak{#1}}}
\newcommand{\bvec}[1][\gamma]{\vec{b}_{\mathfrak{#1}}}
\newcommand{\Tr}{\text{Tr}}
\newcommand{\Vcas}{V_{\text{Cas}}}

\title{
Scale-separated AdS vacua on Riemann-Flat Manifolds  
}
\date{}

\author[a,b]{Miquel Aparici}
\author[c,d]{Ivano Basile}
\author[a,e]{Bruno Valeixo Bento}
\author[a,f]{Miguel Montero}

\affiliation[a]{Instituto de F\'{i}sica Te\'{o}rica IFT-UAM/CSIC,
C/ Nicol\'{a}s Cabrera 13-15, Campus de Cantoblanco, 28049 Madrid, Spain}
\affiliation[b]{Departamento de F\'{i}sica Te\'{o}rica, Universidad Aut\'{o}noma de Madrid, Cantoblanco, 28049 Madrid, Spain}
\affiliation[c]{INFN sezione di Torino, via Pietro Giuria 1, 10125 Torino, Italy}
\affiliation[d]{Jefferson Physical Laboratory, Harvard University, Cambridge, MA 02138, USA}
\affiliation[e]{Laboratoire d'Annecy-le-Vieux de Physique Th\'eorique (LAPTh), CNRS,
Universit\'e Savoie Mont Blanc (USMB),
9 Chemin de Bellevue, 74940 Annecy, France}
\affiliation[f]{CERN, Theoretical Physics Department, 1211 Meyrin, Switzerland}

\emailAdd{miquel.aparici@estudiante.uam.es}
\emailAdd{ivano.basile@to.infn.it}
\emailAdd{bruno.bento@lapth.cnrs.fr}
\emailAdd{miguel.montero@csic.es}

\abstract{Supergravity compactifications supported by fluxes and Casimir energy offer a simple strategy to concoct explicit, controlled and isolated vacua with parametric scale separation and soft supersymmetry breaking. Exploiting the special properties of both the Casimir and flux potentials with respect to Laplacian operators in moduli space, we prove that the simplest such class of solutions, given by toroidal compactifications of eleven-dimensional supergravity threaded by any fluxes, always exhibit perturbative instabilities in the form of tachyonic modes below the BF bound. This no-go theorem applies to any toroidal supergravity compactification supported by $p$-form fluxes.  We also study compactifications on more general Riemann-flat manifolds (RFM's). Concretely, we find a tachyon-free AdS$_4$ vacuum by compactifying M-Theory on a $T^7/\mathbb{Z}_2^3$. \underline{This constitutes the first ever example of a fully top-down, perturbatively stable,} \underline{scale-separated AdS without localized sources.}}

\begin{document}
\emergencystretch 3em
\hypersetup{pageanchor=false}
\makeatletter
\let\old@fpheader\@fpheader
\preprint{IFT-26-148 \par\hfill CERN-TH-2026-258}

\makeatother

\maketitle

\hypersetup{
    pdftitle={},
    pdfauthor={},
    pdfsubject={}
}

\section{Introduction} 
\label{sec:introduction}

One of the pressing questions in holography, string theory, and the Swampland, is the fate of scale-separated vacua (see \cite{Coudarchet:2023mfs} for a review). Supersymmetric scale-separated vacua, like the DGKT construction \cite{DeWolfe:2005uu},  have received the most attention from the community (see e.g. \cite{Farakos:2020phe, Emelin:2022cac, Tringas:2023vzn, Andriot:2023fss, Farakos:2023nms, Cribiori:2024jwq, Tringas:2025bwe, Miao:2025rgf, Tringas:2025uyg, Tringas:2026ptt, Tringas:2026ncg, Montero:2022ghl, Apers:2022zjx, Lust:2022lfc, Bena:2024are, Bedroya:2025ltj, Bedroya:2025fie, Bobev:2025yxp, Revello:2026eqp, Revello:2026opc, Revello:2026yla, Apers:2026lgi, Apers:2026ahu, Cribiori:2026btb, Aggarwal:2026jan} for recent works on the topic) and are notoriously hard to understand. 

Interestingly, non-supersymmetric scale-separated vacua are much easier to tackle! In \cite{Luca:2022inb}, the authors exhibited a simple non-supersymmetric $AdS_4\times T^7$ vacuum of M-theory with $G_7$ flux which is scale separated. If one is interested only in a proof of existence of such vacua, \cite{Luca:2022inb} is enough. But more often we are interested in stable or, at least, perturbatively stable vacua\footnote{At the non-perturbative level, we expect all non-supersymmetric vacua to be unstable \cite{Ooguri:2016pdq, Freivogel:2016qwc}, and this is explicitly the case for all AdS vacua on $T^k$ supported by flux \cite{Aparici:2025kjj}.}. In \cite{Aparici:2025kjj}, two of us showed that the AdS solution of \cite{Luca:2022inb} has tachyons below the Breitenlohner-Freedman (BF) bound \cite{Breitenlohner:1982bm}, and is hence perturbatively unstable. 

However, the analysis in \cite{Aparici:2025kjj} was limited to the original proposal of \cite{Luca:2022inb}, where the $T^7$ was taken at the particular point in moduli space where all sides are equal. Could there be some other point in moduli space where the compactification would be BF-stable? Could, perhaps, a different toroidal compactification with flux (e.g. IIB on $AdS_4\times T^6$ with 5-form flux) be perturbatively stable?

One of the main results of this paper is a general theorem that this is not possible: we show that, breaking supersymmetry by compactification, any $AdS\times T^k$ vacuum of string/M-theory supported by fluxes and Casimir energies has BF tachyons. Notably, our proof applies to \emph{any} point in the moduli space of the parent torus. This is remarkable, since to determine the scalar potential at each point of moduli space we must first compute the Casimir energy. As shown in \cite{ValeixoBento:2025yhz, ValeixoBento:2025qih, ValeixoBento:2026jdx} (see also \cite{DallAgata:2025jii}), the only known way to compute these Casimir energies is via a numerically intensive and laborious algorithm, and in a direct approach to the problem, one would have to do this for every point in moduli space, identify the AdS saddles in the corresponding high-dimensional problem, and show that there are BF tachyons at every point. 

Such a direct approach is clearly unfeasible. Instead, our proof uses a key elementary fact which, to our knowledge, has not been exploited in the literature: That the Casimir energy $\Vcas(x^a)$, as a function of moduli $x^a$, is an eigenfunction of the moduli space Laplacian with a known eigenvalue, for any spin structure and dimension\footnote{Reference \cite{Luo:2022tqy} contains essentially this observation, but restricted to a bosonic free field theory compactified on $T^k$, and \cite{Cheng:2001ti} discussed the $D=2$ case.}. Furthermore, flux terms are also eigenfunctions of the same moduli space operator. Using these properties, it is possible to put bounds on the masses of any AdS vacuum in the moduli space, and show the universal existence of BF tachyons, without ever computing explicitly a Casimir energy or constructing a single concrete solution --- the fact that the different terms in the scalar potential satisfy differential equations in moduli space is all that is needed.  The way in which we learn about properties in the interior of moduli space without actually computing anything there is reminiscent of the fixed-point approach of \cite{Lust:2024aeg}, although in a non-supersymmetric setup. 

The key observation that the scalar potential obeys simple differential equations in moduli space is the first non-supersymmetric manifestation of a pattern observed in many cases for supersymmetric string compactifications, where it has been shown that several quantities of interest satisfy simple differential equations in moduli space: Calabi-Yau periods and Picard-Fuchs equations \cite{Ceresole:1992su}, masses of BPS states in theories with $\mathcal{N}=4$ \cite{Etheredge:2023zjk,Aoufia:2026ztl} and $\mathcal{N}=2$ theories, the 4d $\mathcal{N}=2$ prepotential \cite{Alvarez-Garcia:2026uaq,Alvarez-Garcia:2026vwq}, and even the whole moduli space metric can be obtained as a solution of Einstein's equations in moduli space \cite{Cecotti:2020uek}. See also \cite{Grimm:2020cda,Grimm:2021ikg} for attempts to define these effective field theories from the boundaries via a notion of moduli space holography, and \cite{Aoufia:2024awo, Basile:2024lcz, Aoufia:2026bau} for similar differential equations in the context of worldsheet conformal manifolds. 

The standard reason one gives for these phenomena is that some subset of protected quantities, usually due to supersymmetry, obey a finite-dimensional system of differential equations, and hence obey local equations in moduli space.  One might have been inclined to view this phenomenon as an artifact of supersymmetry. However, the present paper provides a non-supersymmetric avatar of the same idea. In this case, this occurs because the Casimir energy happens to be an eigenfunction of the moduli space Laplacian, but there could be other protected quantities in other interesting examples that can be described by a ``moduli space effective field theory''. Exploring the extent of this phenomenon remains an interesting question for the future.

In the second part of this paper, we extend our study to compactifications on more general Riemann-flat manifolds (RFM's). This allows us to bypass the no-go for the torus, since the  Casimir energy in an RFM no longer obeys a simple Laplace eigenvalue equation.  As an illustration, we obtain a perturbatively stable, parametrically scale-separated family of AdS$_4$ vacua of M-theory. To our knowledge, this is the first example of a fully controlled, scale-separated top-down vacuum without perturbative instabilities. Importantly,  no localized sources are used, unlike in all supersymmetric examples (see \cite{Coudarchet:2023mfs} for a review), which allows us to study the 11d backreaction explicitly and show it is small. The compactification RFM is a quotient of the form  $T^7/\mathbb{Z}_2^3$, where the $\mathbb{Z}_2^3$ holonomy group sits inside $G_2$---importantly though, supersymmetry is broken by an antiperiodic spin structure in the parent $T^7$.  

The AdS vacuum we obtain has no scalars of negative mass squared (nor other tachyonic modes); in that sense, it is one of the few explicit AdS vacua with no scalar close to the BF bound (other examples include some AdS vacua of the $O(16)\times O(16)$ heterotic string \cite{Basile:2018irz, Basile:2026lyc}, and $AdS_4\times S^7/\mathbb{Z}_k$ with $k\geq 6$ and a supersymmetry-breaking spin structure \cite{Bobev:2021wir}). Despite its \emph{perturbative} stability, this vacuum is unstable towards non-perturbative nucleation of M2 branes, as shown in \cite{Aparici:2025kjj}. It nevertheless remains an interesting potential candidate for non-unitary AdS holography, as in \cite{Vafa:2014iua}, or unitary but non-conformal holography, as in \cite{Antonelli:2018qwz, Ghosh:2021lua, Basile:2022zee}.

The rest of the paper is organized as follows. Section \ref{sec:Casimir-compactifications} introduces the AdS compactifications under study, their Casimir energies, and the torus moduli space and its Laplacians. Section \ref{sec:PDE} shows that the Casimir energy on the torus is an eigenfunction of the moduli space Laplacian, and Section \ref{sec:no-go-torus} uses this to prove our main no-go theorem. Section \ref{sec:RFM} discusses the extent to which these ideas can be applied to the more general setting of compactifications on Riemann-flat manifolds, and presents three explicit examples of AdS vacua---the last example is the tachyon-free AdS solution mentioned above, which we discuss in some detail. Finally, Section \ref{sec:conclus} contains our conclusions.

\section*{Notation}
In this paper, we use the following notation and conventions
\begin{itemize}
    \item Mostly plus metric, $(-,+,...,+)$;
    \item We compactify a $D$-dimensional theory down to $d$ dimensions on a $k$-dimensional compact space; the coordinates on the compact space are $\vec{z}$; we use $p$ for $p$-form fluxes.
    \item $\mathbf{g}$ denotes the metric on $T^k$/RFM$_k$ and $\mathbf{g} = R^2\,\hat{\mathbf{g}}$, with $\hat{\mathbf{g}}$ the volume-independent metric on $T^k/$RFM$_k$ with $\det\hat{\mathbf{g}} = 1$; 
    \item $\mathbf{G}$ denotes the metric on the moduli space of metrics, with $\hat{\mathbf{G}}$ the metric on the subspace orthogonal to the volume modulus $R$; we make this choice to make lowercase $g_{ij}$ consistent with the Laplacian definition below.
    \item We use the positive sign convention for the Laplacian, times a factor of 4 \cite{Obers:1999um}
    \begin{equation}
        \Delta_{\mathcal{M}_p} = g_{ik}g_{jl}\frac{\partial}{\partial \tilde g_{ij}}\frac{\partial}{\partial \tilde g_{kl}} + (p+1)g_{ij}\frac{\partial}{\partial \tilde g_{ij}}\,,
    \end{equation}
    so that the Laplacian gives directly the trace of the Hessian. The factor of 4 comes from the kinetic term $\frac{1}{4}\text{Tr} (g^{-1} dg)^2$.
    \end{itemize}

\section{Casimir compactifications}
\label{sec:Casimir-compactifications}
Let us begin with a summary of the basic formulae involved in the Casimir-flux compactifications of interest, and introduce the Laplacian operators in moduli space we will use later.

As noted in the introduction, Casimir compactifications offer a simple setting where to construct parametrically scale separated vacua. The first construction of this type was presented in \cite{Luca:2022inb}. Their solution is based on the low-energy compactification of M theory, eleven-dimensional supergravity. To be precise, they compactified the theory on a square seven-torus and introduced a 7-form flux (dual to the 4-form $F_7 = \star F_4$) threading the internal manifold. A non-vanishing Casimir energy is achieved by imposing antiperiodic boundary conditions for the fermions along every torus cycle. In the end, one obtains a four dimensional effective theory with a scalar potential given by
\begin{equation}\label{eq:4d.potential}
    \frac{V_{4}}{M_{4}^4} = \frac{\ell_{11}^{18}}{(2\pi R)^7} \left (\frac{\ell_{11}^{3}}{2R^{14}}\left ( \frac{N}{2\pi} \right )^2 - \frac{2\mathcal{C}}{R^{11}} \right),
\end{equation}
where $R$ represents the size of the torus cycles, $N$ is the 7-flux number, and $\mathcal{C}$ is a function of the shape moduli. Importantly, the square torus sits at a privileged point in moduli space, where the derivatives of $\mathcal{C}$ with respect to all moduli vanish, and finding a saddle for this potential reduces to extremizing the radius $R$. Concretely, there exists a critical point where the torus and AdS lengths are given respectively by
\begin{equation}\label{eq:radii}
    R  = \ell_{11} \left ( \frac{N}{2\pi}\right )^{\frac{2}{3}} \mathcal{C}^{-\frac{1}{3}} \frac{7^{\frac13}}{2 \cdot 3^{\frac{1}{3}}} \, , \quad
    L  = \ell_{11} \left ( \frac{N}{2\pi}\right )^{\frac{11}{3}} \mathcal{C}^{-\frac{7}{3}} \, \frac{7^{\frac{7}{3}}}{2^{\frac{11}{2}}\cdot 3^{\frac{4}{3}}} \, .
\end{equation}
This solution exhibits parametric scale separation when the flux number $N$ becomes large\footnote{Note that this is a stronger scale separation than, for example, in DGKT \cite{DeWolfe:2005uu}, where $R/L \propto N^{-\frac{1}{2}}$.},
\begin{equation}
    \frac{R}{L } \sim N^{-3} \ll 1 \, .
\end{equation}
In this limit, the radii are parametrically larger than the Planck scale, rendering the supergravity approximation valid, and one can also check that higher derivative and curvature corrections remain under parametric control. Together with scale separation, this allows us to consistently compute the Casimir energy by approximating AdS with flat space \cite{Aparici:2025kjj,ValeixoBento:2025yhz}. In addition, since the energy density due to fluxes is small in Planck units, we can neglect the contribution of fluxes to the fluctuation operators entering the Casimir energy via the trace-log one-loop formula\footnote{In more detail, if the eleven-dimensional energy density due to fluxes is $\rho_\text{flux} \equiv \ell_{11}^{-9}M_\text{flux}^2$ in terms of the mass scale $M_\text{flux}$, fluctuation operators can have terms where derivatives are replaced by $M_\text{flux}$. By parity symmetry, the first correction to the Casimir energy density $\delta \rho_\text{Cas} \sim \ell_\text{comp}^{-9}M_\text{flux}^2$ will depend on some compactification length scale $\ell_\text{comp} \gg \ell_{11}$, and thus will be subleading relative to the classical flux term.}. Similar considerations will apply throughout this paper.

In \cite{Aparici:2025kjj}, it was shown that this family of vacua suffers from perturbative instabilities: the solution does not sit at a true minimum. Specifically, analyzing the behavior of the potential under variations of two torus cycle sizes, while keeping the overall volume fixed, reveals the presence of a tachyonic direction, with a squared mass
\begin{equation}\label{eq:negative.mass}
    m^2_xL^2 = -\frac{k(d-1)^2}{2(k-d)} \frac{\partial^2_x\mathcal{C}}{\mathcal{C}} \approx -342.73 \, ,
\end{equation}
which is well below the \ac{BF} bound \cite{Breitenlohner:1982bm, Breitenlohner:1982jf}. Instabilities of this kind are also present in similar constructions with positive vacuum energy \cite{ValeixoBento:2025yhz}, with negative squared masses that are of the same order of magnitude. One of the main goals of this paper is to assess whether these instabilities are universal in all Casimir compactifications or if, on the contrary, it is possible to construct a parametrically scale separated solution that is perturbatively stable. Whichever the answer to this, non-perturbative instabilities are expected to be present due to brane nucleation \cite{Ooguri:2016pdq}.

In general, the scalar potential that arises from compactifying a $D$-dimensional theory down to $d=D-k$ dimensions, on a compact manifold of dimension $k$, with $p$-form fluxes $(p\leq k)$ turned on along the internal manifold and Casimir energies is given by
\begin{equation}
    \frac{V(R,x^a)}{(M_d)^d} = \frac{1}{R^{\frac{d}{d-2}\,k}}\left(\sum_p \frac{f_p(x^a)}{R^{2p-k}} - \frac{\mathcal{C}(x^a)}{R^d}\right) \,, 
    \label{eq:general-potential}
\end{equation}
where $f_p(x^a) \equiv \mathcal{G}^{(p)}_{\alpha \beta}(x^a)\,n_p^\alpha n^\beta_p$ and $\mathcal{C}(x^a)$ are only functions of the moduli $x^a$ orthogonal to the volume modulus $R$, corresponding to each flux and Casimir contributions respectively. We have also included an overall factor corresponding to the Weyl rescaling necessary to go to Einstein frame in $d$ dimensions, so that the potential is given in $d$-dimensional Planck units. Note that the only dimensionful quantity in the potential \eqref{eq:general-potential} is the volume modulus $R$, with all other quantities being dimensionless. We will remove all dependence in fundamental constants by measuring the radius in units of $\ell_D$, i.e. we work in $D$-dimensional Planck units.  Thus, our solutions will be under control whenever $R$ is large. 
We can write the potential in the simple form 
\begin{equation}
    \frac{V(R,x^a)}{(M_d)^d} = \sum_p \frac{f_p(x^a)}{R^{\gamma_p}} - \frac{\mathcal{C}(x^a)}{R^{\gamma_c}} \,, 
    \label{eq:general-potential-simple}
\end{equation}
by defining the quantities 
\begin{equation}
    \gamma_p = \frac{2 ((d-2)p + k)}{d-2} \,,\quad
    \gamma_c = \frac{d (d+k-2)}{d-2} \,.
    \label{eq:gamma-defs}
\end{equation}
Since the functions $f_p(x^a)$ are always non-negative (they measure volumes in the internal manifold), an AdS saddle point necessarily requires $\mathcal{C}(x^a)>0$ in order to have a negative term in the potential for $R$. In turn, this requires a surplus of bosonic over fermionic contributions to the Casimir energy.  Furthermore, we must also have $\gamma_p > \gamma_c \implies D=d+k < 2p$ for at least one non-vanishing $p$-form flux, so that a positive-decreasing flux contribution dominates before the negating-increasing Casimir term as $R$ increases. This last constraint implies that one cannot have a top-down Casimir AdS minimum with $d\geq 5$. Provided that both conditions are satisfied, a critical point along the $R$ direction is found for 
\begin{equation}\label{eq:radius_eom}
    \sum_p \frac{\gamma_p}{\gamma_c}\,\frac{f_p(x^a)}{R^{\gamma_p}} = \frac{\mathcal{C}(x^a)}{R^{\gamma_c}} \,,
\end{equation}
which reduces to a polynomial equation for $R$; the value of the potential at the critical point is then
\begin{equation}
    V_{\rm crit} = -\sum_p \frac{\gamma_p - \gamma_c}{\gamma_c}\,\frac{f_p(x^a)}{R^{\gamma_p}} \,.
    \label{eq:Vsaddle}
\end{equation}
This is an AdS saddle whenever $V_{\rm crit} < 0$, with an AdS length scale in $d$-dimensional Planck units given by
\begin{equation}
    L^2 = -\frac{(d-1)(d-2)}{2V_{\rm crit}} \,.
    \label{eq:AdS-scale}
\end{equation}
A sufficient condition for $V_{\rm crit}<0$, although not necessary, is that $\gamma_p - \gamma_c = 2p -D > 0$ for all $p$-form fluxes that are turned on. This condition effectively sets a lower bound on $p$ that guarantees an AdS saddle.

Furthermore, the large-flux limit guarantees that any solution will be a scale-separated AdS. To see this, notice that one can recast \cref{eq:radius_eom} as
\begin{equation}\label{eq:radius_eom_recast}
	\sum_p A_p(x^a) \, \left(\frac{1}{R}\right)^{\! 2p-D} = 1.
\end{equation}
where at least one of the positive prefactors $A_p(x^a)$ is large. Therefore, in order for the left-hand side to remain bounded, it must be that $2p>D$ for all $p$ appearing in the sum, and $R$ must be parametrically large in Planck units, ensuring that the solution is under control. Then, on account of \cref{eq:Vsaddle}, the on-shell potential will be negative.

In order for this AdS vacuum to be perturbatively stable, all moduli masses must be above the \ac{BF} bound \cite{Breitenlohner:1982jf,Breitenlohner:1982bm}, 
\begin{equation}
    m_I^2\,L^2 > - \frac{(d-1)^2}{4} \,.
    \label{eq:BF-bound}
\end{equation}
These masses are directly computed from the scalar potential \eqref{eq:general-potential-simple} as the eigenvalues of the mass matrix 
\begin{equation}
    M^I_{\phantom{I}J} = G^{IK}\partial_K\partial_J V \,,
    \label{eq:mass-matrix}
\end{equation}
by taking into account the field space metric $\mathbf{G}$ that guarantees canonically normalised perturbations around the AdS vacuum solution; for metric moduli, $\mathbf{G}$ corresponds to the metric on the moduli space $\M$ of metrics $\mathbf{g}$ defined on the compact space. This moduli space metric can be read off of the moduli kinetic terms that arise upon dimensional reduction from the $D$-dimensional theory, and for compactifications on tori is given by (see e.g. \cite{Etheredge:2022opl})
\begin{equation}
    ds^2_\M = \frac{1}{d-2}\big(d\log\sqrt{\det\,\mathbf{g}}\big)^2 + \frac14\Tr\big[(\mathbf{g}^{-1}\,d\mathbf{g})^2\big] \,.
    \label{eq:moduli-space-metric-def}
\end{equation}
From the point of view of this metric there is a convenient parametrisation of $\mathbf{g}$ that factors out the overall volume dependence as $\mathbf{g} = R^2\,\hat{\mathbf{g}}$, where $\det\hat{\mathbf{g}} = 1$, since for this choice the metric splits neatly into two blocks,
\begin{align}
    ds^2_\M &= \frac{k(d+k-2)}{d-2}\,(d\log\,R)^2 + \frac14\Tr\big[\big(\hat{\mathbf{g}}^{-1}\,d\hat{\mathbf{g}}\big)^2\big] \nonumber \\
    &= \frac{k(d+k-2)}{d-2}\,\frac{dR^2}{R^2} + \hat{G}_{ab}\,dx^a dx^b \,,
    \label{eq:moduli-space-metric}
\end{align}
one for the volume modulus $R$, and another for all remaining moduli $x^a$ that become orthogonal to $R$. For this particular parametrisation, the mass matrix also splits into a volume component and a block that only relates to the shape moduli $x^a$. We can therefore study these fields independently from $R$; in fact, for the volume modulus we find 
\begin{equation}
    m_R^2\,L^2 
    = -\frac{(d-1)^2}{4}\,\frac{2d(d-2)}{k(d-1)}\,\frac{\sum_p \gamma_p\,\frac{\gamma_p - \gamma_c}{\gamma_c}\,\frac{f_p(x^a)}{R^{\gamma_p}}}{V_{\rm crit}} \,, 
    \label{eq:R-mass}
\end{equation}
which is positive whenever the sufficient condition for an AdS vacuum, i.e. $2p>D$, is satisfied.
It follows that the volume direction is always stable in these AdS vacua. An instability can only arise from the $x^a$, and we will therefore focus our attention on these moduli. 

To compute the physical masses of shape moduli at a critical point, it is helpful to introduce the natural differential Laplace operator $\Delta_{\M_k}$ on moduli space. This is simply the natural Laplacian associated to the metric \eqref{eq:moduli-space-metric}, and is given by the expression
\begin{equation}
    \Delta_{\M_k} = \frac{1}{\sqrt{\det\mathbf{G}}}\,\partial_I\big(\sqrt{\det\mathbf{G}} \,G^{IJ}\partial_J\big) 
    = \frac{d-2}{k(d+k-2)}(R\,\partial_R)^2 + \hat{\Delta} \,.
    \label{eq:Laplacian-definition}
\end{equation}
Due to the block structure of $\mathbf{G}$, the Laplacian operator also splits into two separate terms, one corresponding to the volume modulus and another which is the Laplacian defined in the subspace orthogonal to volume deformations, whose induced metric is $\hat{\mathbf{G}}$,
\begin{equation}
    \hat{\Delta} = \frac{1}{\sqrt{\det\hat{\mathbf{G}}}}\,\partial_a\big(\sqrt{\det\hat{\mathbf{G}}} \,\hat{G}^{ab}\partial_b\big) \,.
\end{equation}
 Interestingly, the moduli space operator $\hat{\Delta}$ can be computed in terms of the metric $g_{ij}$ of the physical torus, as
\begin{equation}
    \hat\Delta = g_{ik}g_{j\ell}\frac{\partial}{\partial \tilde g_{ij}}\frac{\partial}{\partial \tilde g_{k\ell}} + (k+1)g_{ij}\frac{\partial}{\partial \tilde g_{ij}}
    - \frac{1}{k}\left(g_{ij}\frac{\partial}{\partial\tilde{g}_{ij}}\right)^2\,.
    \label{eq:explicit.laplacian.unit}
\end{equation}
where $\tilde{g}_{ij}$ is defined such that $\frac{\partial g_{ij}}{\partial \tilde{g}_{k\ell}} = \delta^k_i\delta^\ell_j + \delta^k_j\delta^\ell_i$ \cite{Obers:1999um}. This is how we will compute $\hat{\Delta}$ in practice.

Importantly, when applied to the scalar potential and evaluated at a critical point for which $\partial_I V = 0$, the Laplacian gives exactly the trace of the mass matrix \eqref{eq:mass-matrix},
\begin{equation}
    \Delta_{\M_k}\,V \big|_{\rm saddle} = G^{IJ}\partial_I\partial_J V = \Tr\,\mathbf{M} \,.
\end{equation}
This relation allows us to relate the masses in a saddle to the properties of the total potential $V$ with respect to $\hat{\Delta}$, and is the key for the results presented in the next sections. Restricting ourselves to the trace of the mass matrix on the shape moduli, and replacing $L^2\sim 1/V$ \eqref{eq:AdS-scale}, we obtain
\begin{equation}
    \sum_a m_a^2\,L^2 = -\frac{(d-1)(d-2)}{2}\,\frac{\hat\Delta V}{V}.\end{equation}
    Thus, the sign and magnitude of $\frac{\hat\Delta V}{V}$ informs us about the stability of the vacuum. For a stable vacuum, we must impose the \ac{BF} bound \eqref{eq:BF-bound} on every modulus, leading to
\begin{align}
    \frac{\hat\Delta V}{V} < \frac{d-1}{4}\cdot\frac{(k-1)(k+2)}{d-2} \,.
    \label{eq:LapV-BF-inequality}
\end{align}
This is a necessary condition to have a perturbatively stable AdS vacuum. 
As we will see, for torus compactifications it is possible to use the properties of $\hat\Delta V$ to show not only that \eqref{eq:LapV-BF-inequality} is always violated. 

\section{Casimir energy, fluxes and Laplacian operators}
\label{sec:PDE}
Before proving our no-go theorem for torus compactifications in \secref{sec:no-go-torus}, we will now show that all terms in the scalar potential \eqref{eq:general-potential} satisfy an eigenvalue equation for the Laplacian operators $\Delta_{\M_k}$ and $\hat\Delta$, and therefore correspond to eigenfunctions of these operators with eigenvalues that depend only on the values of $(d,k,p)$. This will make the bound straightforward.
\subsection{Differential equation for the Casimir}
\label{sec:PDE-Casimir}

We will now show that the Casimir energy on a flat $k$-torus, as a function over its moduli space $\mathcal{M}_k$ of flat metrics, is a (non-normalizable) eigenfunction of the Laplacian $\Delta_{\mathcal{M}_k}$. By ``Casimir energy'', we mean the one-loop vacuum energy of the higher-dimensional effective field theory compactified on the $k$-torus, and where the UV-divergent piece corresponding to the higher-dimensional cosmological constant has been dropped. The justification to drop this is that in this paper we consider compactifications of supergravity, or more generally higher-dimensional \acp{EFT} whose cosmological constant vanishes due to supersymmetry. Therefore, the vacuum energy of the lower-dimensional \ac{EFT} is not \ac{UV}-sensitive and coincides with the Casimir energy just defined.

As anticipated in the previous section, it will prove convenient to separate the overall volume modulus from the rest. Compactifying on a torus $T^k \equiv \mathbb{R}^k/\Lambda(R,\hat{\mathbf{g}})$, where $\Lambda(R,\hat{\mathbf{g}}) \simeq \mathbb{Z}^k$ is a lattice encoding a unit-determinant metric $\hat{\mathbf{g}}$ and volume modulus $R$, the Casimir energy arising from the supergravity spectrum can be written as a Schwinger integral of the (supertrace of the) heat kernel on the torus. On the background $\mathbb{R}^d \times T^k$, the result is proportional to
\begin{eqaed}\label{eq:cas_pot}
    \Vcas(R,\hat{\mathbf{g}}) = \int_0^\infty \frac{dt}{t^{\frac{d}{2}+1}} \, \left(Z^{\text{Bose}}_{\Lambda}(t) - Z^{\text{Fermi}}_{\Lambda}(t) \right) ,
\end{eqaed}
where each heat kernel is a lattice sum of the form
\begin{eqaed}\label{eq:ss_Z}
    Z^v_{\Lambda}(t) = \sum_{\ell \in \Lambda + v} e^{-\frac{\ell \cdot \hat{\mathbf{g}}^{-1} \cdot \ell}{R^2} \, t}
\end{eqaed}
with some shift vector $v$ implementing the Scherk-Schwarz twist, i.e. the non-trivial boundary conditions for the higher-dimensional fields. One way to see that $\Vcas$ is an eigenfunction of the moduli-space Laplacian is to integrate \cref{eq:cas_pot} termwise, leading to a combination of $GL(k)$ Eisenstein series \cite{Obers:1999um}. 

However, we can get to the same conclusion from direct algebraic manipulations. To do this, we observe that each summand in the lattice sums of \cref{eq:ss_Z} satisfies a linear partial differential equation; applying the Laplacian operator $\Delta_{\mathcal{M}_k}$ to a term in this sum, one finds
\begin{equation}
    \frac{1}{4}e^{m^i g_{ij} m^{j}t} \Delta_{\mathcal{M}_k} e^{-m^i g_{ij} m^{j}t} = (m^ig_{ij}m^j)^2t^2 - \frac{k+1}{2}(m^ig_{ij}m^j) t \, .
\end{equation}
Since the above combinations can be obtained by suitably differentiating with respect to $t$ as well, one arrives at
\begin{eqaed}\label{eq:pde_Q_new}
    \left(-t^2 \partial_t^2 - \frac{k+1}{2} t \partial_t + \frac{1}{4} \, \Delta_{\mathcal{M}_k} \right) e^{-m^i g_{ij} m^{j}t} = 0 \, .
\end{eqaed}
Since this equation is linear, the lattice sums $Z_\Lambda^v$ also satisfy it. Multiplying the lattice sums by $t^{a}$ modifies \cref{eq:pde_Q_new} according to
\begin{eqaed}\label{eq:dt_term}
    \left(-t^2 \partial_t^2 - \frac{k+1}{2} t \partial_t\right)(t^a Z_\Lambda^v) & = t^a\left(-t^2\partial_t^2 - \frac{4a+k+1}{2} t \partial_t - \frac{a(k-1+2a)}{2}\right) Z_\Lambda^v \\
    & = t^a\left(-2a t \partial_t- \frac{a(k-1+2a)}{2} - \frac{1}{4} \, \Delta_{\mathcal{M}_k} \right) Z_\Lambda^v \\
    & = \left(-2a t \partial_t- \frac{a(k-1-2a)}{2} - \frac{1}{4} \, \Delta_{\mathcal{M}_k} \right) (t^a Z_\Lambda^v) \, ,
\end{eqaed}
namely
\begin{eqaed}\label{eq:dt_term_pde}
    \left(-t^2 \partial_t^2 - \frac{k+1-4a}{2} \,t \partial_t + \frac{a(k-1-2a)}{2} + \frac{1}{4} \, \Delta_{\mathcal{M}_k} \right)(t^a Z_\Lambda^v) = 0 \, .
\end{eqaed}
This partial differential equation can be integrated over the Schwinger proper time $t$ to remove the differential operators $\partial_t$ and recast \cref{eq:dt_term_pde} as a Laplacian equation. To this end, we pick the measure $dt/t^2$ so that the first term integrates to zero. Similarly, the second term which features a single derivative becomes
\begin{eqaed}\label{eq:tdt_integration}
    \int_0^\infty \frac{dt}{t^2} t \partial_t (t^a Z_\Lambda^v) =  \int_0^\infty \frac{dt}{t^2} (t^a Z_\Lambda^v),
\end{eqaed}
upon integrating by parts and using the fact that the integrand vanishes at both boundaries (at least for the relevant choice of $a$). In order to obtain the relevant differential equation for the Casimir potential $V$, we need $2-a = d/2+1$, i.e. $a=1-d/2$. All in all, placing everything on the left-hand side one finds
\begin{equation}\label{eq:v_new}
    \Delta_{\mathcal{M}_k} \Vcas = d(d+k-1)\,\Vcas \,. 
\end{equation}
With these conventions the eigenvalue is always positive, since $d>2$ and $k>0$. As a special case, for $k=1$ the Laplacian $\Delta_{\mathcal{M}_1} = (R\,\partial_R)^2$ \cite{Maloney:2020nni}, so $\Vcas \propto R^{-d}$ as it should; in this case the scaling with $R$ is evident from the definition of $V$ and dimensional analysis. This is true for all $k\geq 1$, since we defined $R$ as the overall volume modulus which is the only dimensionful scale that appears in the calculation.

On the other hand, in terms of the dimensionless quantity $\C(x^a)$ defined as $\Vcas = R^{-d}\,\C$, cf. \eqref{eq:general-potential}, we obtain 
\begin{eqaed}\label{eq:unit-Laplacian-C}
    \hat{\Delta}\,\C 
    = \frac{d(k-1)(d+k)}{k} \,\C \,,
\end{eqaed}
in formal agreement with the Laplacian equations satisfied by $SL(k)$ Eisenstein series of weight $-d/2$ \cite{Obers:1999um}. The eigenvalue is always positive, except for $k=1$ for which it vanishes. \Cref{eq:unit-Laplacian-C} is the main result of this section. In \secref{sec:no-go-torus}, we will use it to prove a no-go theorem for stable Casimir compactifications on a torus.

\subsection{Differential equation for the fluxes}
\label{sec:PDE-fluxes}

In the potential \eqref{eq:general-potential}, the Casimir term is balanced against a flux contribution that provides the parametric scale separation of the candidate toroidal AdS vacua. We will now show that, in torus compactifications, these flux terms are \emph{also} (non-normalizable) eigenfunctions of the Laplacian, with each $p$-form satisfying a Laplace eigenvalue equation in which the eigenvalue depends on $p$.

In order to do that, we first note that the flux potential arising from a $p$-form $F_p$ is proportional to the Hodge norm $\int |F_p|^2$. Since $F_p$ can always be written as
\begin{equation}
    F_p = \sum_I f_I \alpha_I \,,
\end{equation}
where $\{\alpha_I = dx^{i_1}\wedge...\wedge dx^{i_p}\}$ is a basis of $p$-forms on $T^k$, labelled by the multi-index $I=(i_1,\ldots,i_p)$, the flux potential is given by 
\begin{align}
    \int |F_p|^2 = \sum_{I,J} f_I \,\mathcal{G}_{IJ}\, f_J \,,
\end{align}
in terms of the metric
\begin{equation}
    \mathcal{G}_{IJ}=\int \alpha_I \wedge \star \alpha_J = \frac{1}{p!}\frac{1}{R^{2p-k}}\,\epsilon_{i_1\ldots i_p}\epsilon_{j_1\ldots j_p}\,\hat{g}^{i_1 j_1}\ldots\hat{g}^{i_p j_p} \,.
\end{equation}
We have made the dimensions explicit by taking out the overall factor of $R$ and writing $\mathcal{G}_{IJ}$ in terms of the unit determinant metric $\hat{g}_{ij}$ defined in Section \ref{sec:Casimir-compactifications}. By applying the Laplacian \eqref{eq:explicit.laplacian.unit} to the metric $G_{IJ}$, we find that this metric is an eigenfunction with eigenvalue
\begin{equation}
    \hat\Delta\,\mathcal{G}_{IJ} = \frac{2p(k+2)(k-p)}{k}\,\mathcal{G}_{IJ} \,.
\label{lap0}\end{equation}
Equation \eq{lap0} can be proven by using the fact that the metric (and hence the laplacian $\Delta$) on the torus moduli space is invariant under $g\to g^{-1}$, together with
\begin{equation}
 \Delta  (\epsilon_{i_1\ldots i_p}\epsilon_{j_1\ldots j_p}\,{g}^{i_1 j_1}\ldots{g}^{i_p j_p}) = (-2p(p-1)+2p(k+1))\epsilon_{i_1\ldots i_p}\epsilon_{j_1\ldots j_p}\,{g}^{i_1 j_1}\ldots{g}^{i_p j_p}\,.
\end{equation}
The result then follows by removing the volume part of the laplacian $\Delta \to \hat \Delta$, which amounts to subtracting $4p^2/k$ from the eigenvalue.

Since the dependence on the moduli in the flux potential arises exclusively from $\mathcal{G}_{IJ}$, this means that the flux potential is also an eigenfunction of the Laplacian with the same eigenvalue,
\begin{equation}
    \hat\Delta\,f_{p} = \frac{2p(k+2)(k-p)}{k}\,f_{p} \,.
    \label{eq:unit-Laplacian-fp}
\end{equation}
Therefore we conclude that each $p$-flux term in the potential is also an eigenfunction of the Laplacian $\hat\Delta$ with an eigenvalue that depends on $p$. In the next section we will use this result to show that any toroidal compactification with a single type of $p$-form flux and Casimir energy that gives an AdS vacuum necessarily has at least one field with a mass below the BF bound---in other words, all such AdS vacua are perturbatively unstable.

\section{No-go theorem for torus compactifications}
\label{sec:no-go-torus}

We are now ready to show that all AdS vacua supported by a single type of $p$-form flux and Casimir energy contain at least one tachyon whose mass is below the \ac{BF} bound. As a direct consequence of the previous sections, we have 
\begin{equation}
    \hat\Delta\,V = \sum_p \lambda_p\,\frac{f_p(x^a)}{R^{\gamma_p}} - \lambda_c\,\frac{\C(x^a)}{R^{\gamma_c}} \,,
    \label{eq:unit-Laplacian-V}
\end{equation}
where $\lambda_p$ and $\lambda_c$ are defined through \eqref{eq:unit-Laplacian-fp} and \eqref{eq:unit-Laplacian-C}, respectively, as
\begin{align}
    \lambda_p = \frac{2p(k+2)(k-p)}{k} \,,
    \quad \lambda_c = \frac{d(k-1)(d+k)}{k} \,. 
\end{align}
As in Section \ref{sec:Casimir-compactifications}, when there are several types of $p$-form flux, both $V$ and $\hat\Delta\,V$ at the critical point are sums over $p$ that contain different powers of $R$,
\begin{equation}
    \frac{\hat\Delta\,V}{V} = \frac{\sum_p (\lambda_c\gamma_p - \lambda_p\gamma_c)\,V_p}{\sum_p (\gamma_p - \gamma_c)\,V_p} \,,
    \quad\text{with}\quad 
    V_p\equiv \frac{f_p(x^a)}{R^{\gamma_p}} > 0 \,,
    \label{eq:DV-on-V-general-flux}
\end{equation}
which in the case of a single type of flux is simply
\begin{equation}
    \frac{\hat\Delta\,V}{V} = \frac{\lambda_c\,\gamma_p - \lambda_p\,\gamma_c}{2p-k-d} \,.
    \label{eq:DV-on-V-single-flux}
\end{equation}
This is only a function of the values $(d,k,p)$, so that the inequality \eqref{eq:LapV-BF-inequality} reduces to
\begin{equation}
    \frac{2d(d-1)}{k}\cdot\frac{p(k+2)(k-p)(d+k-2) - (k-1)(d+k)(p(d-2)+k)}{(d+k-2p)(k+2)(k-1)} <  \frac{(d-1)^2}{4}
\end{equation}
and any compactification that violates it is guaranteed to have at least one modulus with a mass below the \ac{BF} bound. With the constraints that the $p$-form flux fits the compact space $(p\leq k)$, that there is a minimum along the volume direction ($2p-k>d$, see Section \ref{sec:Casimir-compactifications}), and only considering compactifications with $k>1$\footnote{Note that if $k=1$ the only modulus is the overall volume and therefore there are no orthogonal shape moduli to analyse.} to $d>2$ dimensions from supergravity theories $(d+k\leq 11)$, only one case survives: a compactification with $d=3$, $k=8$ and $p=6$. This corresponds to an M-theory compactification on $T^8$, but it requires a 6-form flux that does not exist in M-theory. Therefore, all toroidal vacua supported by a single flux and Casimir energy have at least one unstable direction in moduli space. 

In particular, we find the vacuum in \cite{Luca:2022inb} is unstable, in agreement with \cite{Aparici:2025kjj}. The original analysis covered only a square $T^7$; the methods we use here, via differential equations in moduli space, are much more powerful, and they allow us to exclude any other point in the $T^7$ moduli space at large volume. 

We can also extend our results to toroidal vacua supported by multiple fluxes. In this more general case, one can rewrite \eqref{eq:DV-on-V-general-flux} as\footnote{Since $\lambda_p,V_p>0$ for all $p$, the sum $\sum_p\,\lambda_p\,V_p \leq \max_p\{\lambda_p\}\sum_p V_p$ is bounded from above; whenever $\max_p\{\lambda_p\} - \lambda_c \leq 0$, the sum $\sum_p(\lambda_p-\lambda_c)\,V_p \leq 0$, which once divided by the negative $V$ yields a non-negative contribution to \eqref{eq:DV-on-V-general-flux-bound}.}
\begin{equation}
    \frac{\hat\Delta\,V}{V} 
    = \lambda_c + \frac{\sum_p(\lambda_p - \lambda_c)\,V_p}{V} > \lambda_c,
    \label{eq:DV-on-V-general-flux-bound}
\end{equation}
unless $d=3$ and $k=6,7,8$, since in all other cases $\max_p\{\lambda_p\} - \lambda_c < 0$, and we always have $V < 0$ at an AdS critical point. Consequently, combining \eqref{eq:DV-on-V-general-flux-bound} with \eqref{eq:LapV-BF-inequality}, we conclude that any such case will necessarily have a modulus with a mass below the \ac{BF} bound.  

Nonetheless, three cases remain undecided: Compactification to $d=3$ dimensions on a $k$-dimensional torus $T^k$, for $k=6,7,8$. Two of these cases can be also ruled out with a refined version of the argument that led to \eqref{eq:DV-on-V-general-flux-bound}, as we will now see. 

To construct this refinement, we begin with the observation  that the bound \eqref{eq:DV-on-V-general-flux-bound} exploits the fact that $\hat\Delta V$ takes the form
\begin{equation}
    \hat\Delta\,V = \lambda_c\,V + \sum_p(\lambda_p - \lambda_c)\,V_p .
\end{equation}
Even though we derived this relation for $\hat\Delta$, a similar relation can be derived for any other differential operator $\tilde{\Delta}$ in moduli space for which each term of the potential is an eigenfunction. In that case, we can write
\begin{equation}
    \tilde\Delta\,V = \tilde\lambda_c\,V + \sum_p(\tilde\lambda_p - \tilde\lambda_c)\,V_p \,,
\end{equation}
and we can then consider the combined Laplacian equation
\begin{equation}
    \hat\Delta\,V + \tilde\Delta\,V = (\lambda_c + \tilde\lambda_c)\,V + \sum_p(\lambda_p + \tilde\lambda_p - \lambda_c - \tilde\lambda_c)\,V_p \,.
\end{equation}
Since $V$ is negative, the lower bound \eqref{eq:DV-on-V-general-flux-bound} follows from an upper bound on $\hat\Delta V$. If $\tilde\Delta\,V > 0$, then $\tilde\Delta\,V/V<0$, so we can write the inequality
\begin{equation}
    \frac{\hat\Delta\,V}{V} > \frac{\hat\Delta\,V}{V} + \frac{\tilde\Delta\,V}{V}
    = (\lambda_c + \tilde\lambda_c) + \frac{\sum_p(\lambda_p + \tilde\lambda_p - \lambda_c - \tilde\lambda_c)\,V_p}{V} > \lambda_c + \tilde\lambda_c 
    \label{eq:DV-on-V-general-flux-bound-tilde}
\end{equation}
whenever $\max_p\{\lambda_p + \tilde\lambda_p\} - \lambda_c - \tilde\lambda_c \leq 0$. 

Although finding a suitable operator that has all terms in the potential as eigenfunctions is in general not trivial, we already know one such operator---the volume modulus Laplacian. Furthermore, we concluded in \eqref{eq:R-mass} that under the sufficient condition $2p>D$ that guarantees the saddle is AdS, this operator gives a positive contribution to the total Laplacian. Therefore, we will take 
\begin{equation}
    \tilde\Delta \equiv \frac{1}{\chi}\,(R\,\partial_R)^2 \,,
\end{equation}
for some $\chi>0$; then $\tilde\Delta V > 0$ whenever $2p > D$ for all turned on $p$-form fluxes, and $\tilde\lambda_p = \gamma_p^2/\chi$, $\tilde\lambda_c = \gamma_c^2/\chi$ as defined in \eqref{eq:gamma-defs}. 
For this choice of operator, $\max\{\lambda_p + \tilde\lambda_p\} - \lambda_c - \tilde\lambda_c \leq 0$ for $k=6,7$ when $\chi\in(10,17)$. Thus, combining \eqref{eq:DV-on-V-general-flux-bound-tilde} and \eqref{eq:LapV-BF-inequality}, we can also show that $k=6$ and $k=7$ also contain tachyonic instabilities. 

Unfortunately, there is no choice $\chi$ that allows this argument to be applied to the last remaining case of $d=3, k=8$; this leads us once again to an M-theory compactification on $T^8$. This last case can also be shown to be tachyonic by direct analysis:  Since we ruled out vacua with a single flux, both $G_4$ and $G_7$ fluxes must be present.  However, as shown around \cref{eq:radius_eom_recast}, there is no scale-separated controlled solution in this case. Actually, this argument also excludes the cases with $k=1$ and $d>2$, since none of the available fluxes satisfies $2p>d+k$. 

\section{Compactifications on Riemann-flat manifolds}\label{sec:RFM}

A natural generalisation of the toroidal compactifications studied in the previous sections is to consider the broader class of Riemann-flat manifolds (RFM's), which includes many more spaces beyond tori. Not only do RFM's share many of the nice properties of tori, including control under corrections and computable Casimir energies, they may even enhance some of them. Yet, they are also different enough from tori to evade the no-go theorem proved in the previous section---indeed, in Section \ref{sec:stable.ads}, we present a specific RFM compactification that gives rise to a tachyon-free AdS vacuum. 

Let us start by summarising the key properties of RFM's, referring the reader to \cite{ValeixoBento:2025yhz} for more details (see also \cite{borwein2013lattice}); we will then describe what can be said about tachyonic instabilities in the more general setup of flux compactifications on RFM's. All RFM's are a quotient of $\mathbb{R}^k$ with its standard Euclidean metric, by a subgroup $\mathcal{B}\subset ISO(k)$ of the Euclidean group in $k$ dimensions, 
\begin{equation}
    \mathcal{F}_k=\mathbb{R}^k/\mathcal{B} \,,
    \label{rfmdef}
\end{equation} 
whose group action is free, meaning that no $b\in\mathcal{B}$ has fixed points---the quotient is a manifold, rather than an orbifold. The quotient \eqref{rfmdef} can be taken in two steps, where we first quotient $\mathbb{R}^k$ by a lattice $\Lambda\subset\mathbb{R}^k$ generating the translation subgroup to define a torus,
\begin{equation} 
    T^k=\mathbb{R}^k/\Lambda \,,
\end{equation}
and then quotient the torus by $\Gamma$, defined as the quotient of $\mathcal{B}$ by the translation subgroup,
\begin{equation} 
    \Gamma\equiv \mathcal{B}/\sim\,, \quad b\sim b' \quad\text{iff}\quad b'\circ b^{-1}\in \Lambda \,.
\end{equation}
In order for the quotient to be non-empty, elements in $\Gamma$ must preserve the lattice $\Lambda$, meaning that $\Gamma$ is a \emph{finite} group of isometries of $T^k$. The group $\Gamma$ is also called the point group or holonomy group of the RFM; in fact, RFM's are the only manifolds that have finite holonomy groups \cite{eee57c5f-3af2-3ab0-b3c0-3d30ad3fad43}. In practice, representatives of an element $\gamma\in\Gamma$ are given by affine transformations of the form $\vec{z}\to\D\,\vec{z}+\bvec$ where $\bvec$ is taken modulo translations and $\D$ is orthogonal. Demanding that the $\Gamma$ action has no fixed points, so that $\mathcal{B}$ is torsion-free, leads to constraints on the pairs $(\mathbf{D}_\gamma,\vec{b}_\gamma)$. For instance, the equation 
\begin{equation} 
    (\mathbf{D}_\gamma-\mathbf{I})\, \vec{z}+\vec{b}_\gamma=0\,\, \text{mod}\,\Lambda \,.
    \label{fpfcond}
\end{equation}
must have no solution, which requires $\vec{b}_\gamma$ to have a non-zero component along the invariant subspace of $\mathbf{D}_\gamma$ and thus $\mathbf{D}_\gamma$ must have a non-trivial invariant subspace.

Although the Landscape of string compactifications in RFM's is large, it is finite \cite{eee57c5f-3af2-3ab0-b3c0-3d30ad3fad43}, which aligns well with expected finiteness of the Quantum Gravity Landscape \cite{Heckman:2019bzm,Hamada:2021yxy,Delgado:2024skw}. The classification program for RFM's has been completed in dimensions up to six \cite{Cid2001}, with dimensions 7 and higher being hindered by the fact that the classification of crystallographic groups themselves is an open problem \cite{Moeck_2019}. M and F-theory compactifications to four dimensions require seven and eight-dimensional manifolds, respectively, so a classification would be highly beneficial for a detailed exploration of this corner of the string Landscape.

The moduli space of metrics $\mathcal{M}_k$ of an RFM compactification corresponds to a restriction of the moduli space of its covering torus. The allowed metrics on the RFM are those metrics $\mathbf{g}$ on $T^k$ that satisfy \cite{ValeixoBento:2025yhz}
\begin{equation}\label{eq:moduli.fixing}
    \D^T\cdot\mathbf{g}\cdot\D = \mathbf{g} \,, \quad \forall\,\gamma\in\Gamma \,. 
\end{equation} 
The classical metric $\mathbf{G}$ on $\M_k$ is then the same as the one defined on the moduli space of torus metrics \eqref{eq:moduli-space-metric-def}, except for the fact that $\mathbf{g}$ must now satisfy \eqref{eq:moduli.fixing}. 

\subsection{A Laplace (in)equality for RFMs?}
\label{sec:inequalityRFM}
In \secref{sec:PDE} we have shown that the Casimir energy and the flux potential on torus compactifications are eigenfunctions of the Laplacian on the moduli space of torus metrics, which allowed us in \secref{sec:no-go-torus} to put bounds on the moduli masses and show that a wide class of scale-separated AdS vacua on $T^k$ has tachyons below the BF bound. We now want to consider an extension of these results to the case of RFMs. We are therefore interested in the Laplacian operator $\Delta_{\M_k}$ defined with respect to the moduli space metric $\mathbf{G}$ for RFMs, and how it acts on the Casimir and flux potentials that are functions of the RFM moduli. 

To make our analysis more manageable, we will restrict to RFM's with abelian holonomy group. Consider one such RFM with abelian holonomy and let $\{\D[g]\}$ be a set of generators of the (point) group. The constraint \eqref{eq:moduli.fixing} implies that the metric commutes with all group elements
\begin{equation}
     [\D,\mathbf{g}] = 0\,,
\end{equation}
and, consequently, the $\D$ and $\mathbf{g}$ can be simultaneously diagonalized. Since the $\D$ are orthogonal, their eigenvalues are of the form $\pm 1$ or $e^{\pm i\theta}$, with $\theta \in (0,\pi)$, and therefore their real Jordan form consists of two diagonal blocks with entries $\pm1$, together with $2\times2$ rotation blocks corresponding to angles $\theta\neq \pi$,
\begin{align}
    \D = \begin{pmatrix}
        \cos\theta_1 &-\sin\theta_1 & & &  \\
        \sin\theta_1 &\cos\theta_1 & & & \\
        & &\ddots & & \\
        & & &\cos\theta_\ell &-\sin\theta_\ell \\
        & & &\sin\theta_\ell &\cos\theta_\ell \\ 
        & & & & & -1 &\cdots & 0 & & & \\
        & & & & & \vdots &\ddots &\vdots & & & \\
        & & & & & 0 & \cdots & -1 & & & \\
        & & & & & & & & 1 &\cdots & 0 \\
        & & & & & & & & \vdots &\ddots &\vdots \\
        & & & & & & & & 0 &\cdots & 1 \\
    \end{pmatrix} \,.
\end{align}
Since $\D$ and $\mathbf{g}$ should commute in any basis, in the Jordan basis of the $\D$ the metric $\mathbf{g}$ is also block diagonal. 
A rotation block with angle $\theta \neq \pi$ in any of the $\D$ forces the corresponding $2\times 2$ block in the metric to be proportional to the $2\times 2$ identity. We denote the number of such blocks by $\ell_1$. The remaining blocks of the metric in the Jordan basis are unconstrained. The precise number of such blocks, which we denote by $\ell_2$, depends on the RFM under consideration. For example, for a cyclic RFM there can be at most two such blocks, corresponding to the $+1$ and $-1$ blocks in the Jordan form of the generator. All in all, the generically non-vanishing components of the metric in this preferred basis are
\begin{align}\label{eq:RFM.metric}
    \mathbf{g} = \begin{pmatrix}
        g^{(1)} & 0 & & & &  \\
        0 &g^{(1)} & & & & \\
        & & \ddots & & & \\
        & & &g^{(\ell_1)} &0 & \\
        & & &0 &g^{(\ell_1)} & \\ 
        & & & & & \mathbf{g}^{1} & &\\
        & & & & & &\ddots & \\
        & & & & & & & \mathbf{g}^{\ell_2} \\
    \end{pmatrix} \,.
\end{align}
Once the metric is brought into this form, the induced moduli space Laplacian simplifies significantly. Indeed, the kinetic terms of the RFM moduli decompose into a sum over the metric blocks $\mathbf{g}^{(a)}$
\begin{equation}
    \frac14\Tr\big[(\mathbf{g}^{-1}\,d\mathbf{g})^2\big] =\sum_{a}\frac14\Tr\big[((\mathbf{g}^{(a)})^{-1}\,d\mathbf{g}^{(a)})^2\big] \,,
\end{equation}
with the Laplacian orthogonal to the volume splitting accordingly as
\begin{equation}\label{eq:laplacian}
     \hat \Delta = \sum_{a} \Delta_{(a)} - \frac1k\bigg(\sum_b g^{(b)}_{ij}\frac{\partial}{\partial\,g^{(b)}_{ij}}\bigg)^2\,,
\end{equation}
where $\Delta_{(a)}$ is the Laplacian associated with the corresponding block $\mathbf{g}^{(a)}$. For the unconstrained blocks, this is just the Laplacian on a $k_a$--dimensional torus,
\begin{equation}\label{eq:lap.normal}
    \Delta_{(a)}=g^{(a)}_{ik}g^{(a)}_{jl}\frac{\partial}{\partial \tilde g^{(a)}_{ij}}\frac{\partial}{\partial \tilde g^{(a)}_{kl}} + (k_a+1)g^{(a)}_{ij}\frac{\partial}{\partial \tilde g_{ij}^{(a)}}\,,
\end{equation}
with $k_a$ the size of the corresponding block, while for metric blocks associated with rotations, i.e. $2\times2$ blocks that are proportional to the identity and have a single modulus $g^{(a)}$, the Laplacian takes the form
\begin{equation}\label{eq:lap.rotation}
    \Delta_{(a)} = 2\left(g^{(a)}\frac{\partial}{\partial g^{(a)}}\right)^2 \,.
\end{equation}
Finally, we denote by $A$ the set of indices $a$ such that the Laplacian of the block $\mathbf{g}^{(a)}$ takes this form (i.e. corresponding to the metric $2\times 2$ blocks proportional to the identity).

The Laplacian \eqref{eq:laplacian} acts on any function defined over the moduli space $\mathcal{M}_k$ of RFM metrics, which includes in particular Casimir energies computed on these backgrounds \cite{ValeixoBento:2025yhz}
\begin{equation}
    \Vcas = \sum_{\gamma\in\Gamma} \sum_{\mathbf{r}}{\text{Tr}_\mathbf{r}(\D)}\,\mathcal{E}(\gamma)
    \,,\quad
    \mathcal{E}(\gamma) = -\hat{\delta}_{\vec{h}}\,\frac{\Gamma(s_\gamma)}{2\pi^{s_\gamma}}\cdot
    \frac{\sqrt{\det\mathbf{g}_\parallel}}{\vert\Gamma\vert} \sum_{\vec{\xi}\in\,\Xi_\gamma} \frac{e^{2\pi i \,\vec{\beta_\gamma}\cdot\vec{\xi}}}{|\vec{\xi} + \bvec^\parallel|^{2s_\gamma}_\parallel} \,.
    \label{eq:Casimir-RFMs}
\end{equation}
The Casimir energy is given as a sum---over the massless spectrum representations $\bf r$ of the higher-dimensional theory and over all group elements $\gamma\in\Gamma$---of a definite contribution $\mathcal{E}(\gamma)$ that can be computed as an infinite lattice sum specific to each element $\gamma$. This sum is over the lattice $\Xi_\gamma$ corresponding to the projection of $\Z^k$ onto the invariant subspace of $\D$ (of dimension $k'$) and depends on $s_\gamma = \frac{d+k'}{2}$, as well as the vector $\vec{\beta}_\gamma^{\,\parallel}$ encoding the projection onto this subspace of the covering torus spin structure $\vec{h}$ and the projected metric $\mathbf{g}_\parallel$. Interestingly, $\mathcal{E}(\gamma)$ behaves exactly as the contribution to the effective potential that would come from an effective ``Casimir brane'' wrapped on the invariant subspace of $\D$, of tension given by \eqref{eq:Casimir-RFMs} \cite{ValeixoBento:2025yhz}.

A natural question to ask is whether the Casimir energy \eqref{eq:Casimir-RFMs} is still an eigenfunction of the moduli space Laplacian of an RFM. As we show in appendix \ref{ap:laplace.inequality}, this is not the case. Consequently, we cannot prove a no-go for RFMs. Ultimately, the difficulty in finding a proof for RFMs analogous to that of the torus arises for three key reasons. First, because some moduli of the torus are projected out by the action of the holonomy group of the RFM, which may include the would-be tachyon guaranteed by the torus theorem. In Appendix \ref{ap:geometric.RFM}, we offer a geometric perspective on this fact. Secondly, the sign of $\mathcal{E}(\gamma)$ depends on the phases imposed by the boundary conditions of each field in the spectrum, and might be different for different elements $\gamma$ in the sum. Finally, the total Casimir potential for RFMs is a sum over $\gamma\in\Gamma$ weighted by the traces of $\D$ under the representations of the $D$-dimensional theory, which can again result in relative signs between different terms in $\Vcas$. In any case, such a proof could not hold in general, since at least one vacuum solution without tachyons can be constructed, as we will see in Section \ref{sec:stable.ads}.

Nevertheless, note that one need not worry about relative signs in the Casimir potential when the parent theory is maximal supergravity. This is due to a clean factorization property of their spectrum, as we now describe. 
Consider M-theory, whose spectrum comprises the graviton \textbf{44}, the 3-form field \textbf{84} and the gravitino \textbf{128}, labelled by their representation under the 11-dimensional massless little group $SO(9)$. When compactifying this to $d\geq 3$ dimensions, the massless fields decompose into representations of $SO(11-d)\times SO(d-2)$. This is always a subgroup of $SO(8)$ for $d\geq 3$, which we will use to our advantage. One can decompose these $SO(9)$ representations in terms of their $SO(8)$ branching rules \cite{Slansky:1981yr},
\begin{eqaed}
    \mathbf{44}    &= \mathbf{1} \oplus \mathbf{8_v} \oplus \mathbf{35_v}\,, \quad \mathbf{84} = \mathbf{56_v} \oplus \mathbf{28}\,, \quad \mathbf{128} = \mathbf{56_s}\oplus\mathbf{8_s}\oplus\mathbf{56_c}\oplus\mathbf{8_c}\,.
\end{eqaed}
Further, we will also use that
\begin{align}
    &&& \mathbf{8_v}\otimes \mathbf{8_v} = \mathbf{35_v} \oplus \mathbf{28} \oplus \mathbf{1}\,, 
    \quad 
    &&\mathbf{8_s}\otimes \mathbf{8_c} = \mathbf{56_v} \oplus \mathbf{8_v}\,, && \\ 
    &&& \mathbf{8_s}\otimes \mathbf{8_v} = \mathbf{56_c} \oplus \mathbf{8_c}\,, 
    \quad  
    &&\mathbf{8_v}\otimes \mathbf{8_c} = \mathbf{56_s} \oplus \mathbf{8_s}\,. &&
\end{align}
These relations imply the following identities for the traces
\begin{align}
    &\text{Tr}_{\mathbf{44}}(\D)+\text{Tr}_{\mathbf{84}}(\D)-\text{Tr}_{\mathbf{128}}(\D) = \nonumber \\
    &= \Big[\text{Tr}_{\mathbf{1}}(\D)+\text{Tr}_{\mathbf{28}}(\D)+\text{Tr}_{\mathbf{35_v}}(\D)\Big] + \Big[\text{Tr}_{\mathbf{56_v}}(\D) + \text{Tr}_{\mathbf{8_v}}(\D)\Big] \nonumber \\ 
    &\quad- \big[\text{Tr}_{\mathbf{56_c}}(\D)+\text{Tr}_{\mathbf{8_c}}(\D)\Big] -\Big[\text{Tr}_{\mathbf{56_s}}(\D)+\text{Tr}_{\mathbf{8_s}}(\D)\Big] \nonumber \\ 
    &= \big[\text{Tr}_{\mathbf{8_v}}(\D)\big]^2 + \text{Tr}_{\mathbf{8_s}}(\D)\cdot\text{Tr}_{\mathbf{8_c}}(\D) - \text{Tr}_{\mathbf{8_v}}(\D)\cdot\text{Tr}_{\mathbf{8_s}}(\D)-\text{Tr}_{\mathbf{8_v}}(\D)\cdot\text{Tr}_{\mathbf{8_c}}(\D) \nonumber \\
    &= \Big[\text{Tr}_{\mathbf{8_v}}(\D)-\text{Tr}_{\mathbf{8_s}}(\D)\Big]\cdot\Big[\text{Tr}_{\mathbf{8_v}}(\D)-\text{Tr}_{\mathbf{8_c}}(\D)\Big] \,. 
\end{align}
This is reminiscent, and not unrelated, to the type IIA string theory partition function that we would get by compactifying on an extra spatial circle. We can now make use of the following identity \cite{lawson2016spin}
\begin{equation}
    \text{Tr}_{\mathbf{8_s}}(\D)-\text{Tr}_{\mathbf{8_c}}(\D) = \prod_j (e^{i\theta_j/2}-e^{-i\theta_j/2}) = 16 \prod_j \sin \frac{\theta_j}{2} \,,
\end{equation}
where $\theta_j$ are the Cartan angles of $\D$. In an RFM, each element has a non-trivial invariant subspace within the internal manifold, and the holonomy group acts trivially on the non-compact directions. Hence, the subspace fixed by $\D$ is at least $(d-1)$-dimensional. This implies that for $d\geq 3$ at least one of the angles must be zero and $\text{Tr}_{\mathbf{8_s}}(\D)=\text{Tr}_{\mathbf{8_c}}(\D)$. As a direct consequence, we find
\begin{equation}
    \text{Tr}_{\mathbf{44}}(\D)+\text{Tr}_{\mathbf{84}}(\D)-\text{Tr}_{\mathbf{128}}(\D) = \big[\text{Tr}_{\mathbf{8_v}}(\D)-\text{Tr}_{\mathbf{8_c}}(\D)\big]^2 \geq 0 \,.
\end{equation}
Similarly,
\begin{equation}
    \text{Tr}_{\mathbf{44}}(\D)+\text{Tr}_{\mathbf{84}}(\D)+\text{Tr}_{\mathbf{128}}(\D) = \big[\text{Tr}_{\mathbf{8_v}}(\D)+\text{Tr}_{\mathbf{8_c}}(\D)\big]^2 \geq 0 \,,
\end{equation}
and combining the two inequalities, one finds $\text{Tr}_{\mathbf{44}}(\D)+\text{Tr}_{\mathbf{84}}(\D)\geq 0$. Defining the total bosonic and fermionic contributions, $\text{Tr}_{\mathbf{B}}(\D)=\text{Tr}_{\mathbf{44}}(\D)+\text{Tr}_{\mathbf{84}}(\D)$ and $\text{Tr}_{\mathbf{F}}(\D)=\text{Tr}_{\mathbf{128}}(\D)$, respectively, we can argue
\begin{align}
    V_\gamma &= \Big[{\text{Tr}_\mathbf{B}(\D)}\,\mathcal{E}_\mathbf{B}(\gamma)-{\text{Tr}_\mathbf{F}(\D)}\,\mathcal{E}_\mathbf{F}(\gamma)\Big] 
    \leq \Big[{\text{Tr}_\mathbf{B}(\D)}\,-{\left|\text{Tr}_\mathbf{F}(\D)\right|}\Big]\,\mathcal{E}_\mathbf{B}(\gamma) \leq 0 
    \,.
\end{align}
where we have used that $\mathcal{E}_\mathbf{F}(\gamma)\geq \mathcal{E}_\mathbf{B}(\gamma)$ and $\mathcal{E}_\mathbf{B}(\gamma)\leq 0$, which follows directly from the fact that only non-trivial phases can introduce signs in the lattice sums of \eqref{eq:Casimir-RFMs} and that these signs can only make the sums smaller in absolute value. In short, we have argued that the contributions from twisted sectors to the Casimir energy are always negative, no matter what spin structure is chosen for the RFM.
This argument applies to eleven-dimensional supergravity and, by toroidal reduction, to complete ungauged maximal-supergravity multiplets in $D\leq9$, provided the RFM action lifts trivially along the auxiliary torus. It does not automatically cover ten-dimensional type IIB with duality bundles or gauged/truncated maximal theories.


\subsection{Casimir compactifications on RFMs: examples and general considerations} 

In the absence of a proof that RFM flux compactifications with Casimir energy must have BF tachyons, one must consider the possibility that perturbatively stable examples exist in the RFM Landscape---and indeed they do. In this section, we look at some concrete examples, namely an $AdS_3\times T^4/\Z_2$ solution of 7d maximal supergravity, and an $AdS_4\times T^7/\Z_8$ solution in M-theory, using RFM's with cyclic holonomy. Although both have BF tachyons, they present some remarkable features that, put together, allow us to construct a perturbatively stable solution, which we present in Section \ref{sec:stable.ads}. 

In all cases, we consider for simplicity a space-filling flux along the compact directions, so that all moduli dependence (apart from the volume) is contained in the Casimir potential, that is,
\begin{equation} 
    \frac{V_{d}(R,x)}{M_d^d} = \frac{\ell_{d+k}^{d+2k}}{(2\pi R)^k}\left(\frac{\ell_{d+k}^{k-d}}{2R^{2k}}\left ( \frac{N}{2\pi} \right )^2 - \frac{ \mathcal{C}(x)}{R^{d+k}} \right)\, .
\end{equation}
In order to explicitly compute $\mathcal{C}(x)$ and study the critical points, we employ the methods developed in \cite{ValeixoBento:2025yhz} and implemented in \cite{ValeixoBento:2026jdx}.

At the end of this section, we present a heuristic argument showing that a perturbatively stable solution should be expected to have vanishing twisted terms; this will indeed be the case for the solution in Section \ref{sec:stable.ads}.

\subsubsection*{Example 1: \texorpdfstring{$AdS_3 \times T^4/\mathbb{Z}_2$}{AdS3xT4/Z2} in 7d supergravity}

Consider the compactification of 7d maximal supergravity on the four-dimensional cyclic RFM generated by
\begin{equation}
    \D[g]=\left(\begin{array}{*4{C{1.1em}}}
            -1 & 0 & 0 & 0 \\
            0 & -1 & 0 & 0 \\
            0 & 0 & 1 & 0 \\
            0 & 0 & 0 & 1
        \end{array}\right)\,,
        \quad
    \bvec[g] =\begin{pmatrix}
                0 \\
                0 \\
                \frac{1}{2} \\
                \frac{1}{2}
            \end{pmatrix} \,.
\end{equation}
The massless spectrum consists of 1 graviton, 5 two-forms, 10 vector fields, 14 real scalars, 4 gravitinos, and 16 dilatino spinors; 
while all bosonic boundary conditions are periodic, for fermions we will choose antiperiodic boundary conditions along the first two circles, and periodic along the base, i.e. we pick a spin structure on the covering $T^4$ encoded by the vector $\vec{h}=(\tfrac12,\tfrac12,0,0)$. The group action fixes four out of the ten moduli of the torus, leaving six allowed deformations, which we parametrize as
\begin{equation}
    \mathbf{g} = R^2
    \begin{pmatrix}
        e^{\alpha/\sqrt{2}}e^{x_1} \cosh{y_1} & e^{\alpha/\sqrt{2}}\sinh{y_1} & 0 & 0 \\
        e^{\alpha/\sqrt{2}}\sinh{y_1} & e^{\alpha/\sqrt{2}}e^{-x_1}\cosh{y_1} & 0 & 0 \\
        0 & 0 & e^{-\alpha/\sqrt{2}}e^{x_2}\cosh{y_2} &  e^{-\alpha/\sqrt{2}}\sinh{y_2} \\
        0 & 0 &  e^{-\alpha/\sqrt{2}}\sinh{y_2} &e^{-\alpha/\sqrt{2}} e^{-x_2} \cosh{y_2} 
    \end{pmatrix} \,.
\end{equation}
Rather than extremise the potential with respect to all 5 moduli, we can use symmetries in order to determine a critical point. For example, the transformation
\begin{equation}
    A = \left(\begin{array}{*4{C{1.1em}}}
    0 & -1 & 0 & 0 \\
    1 & 0 & 0 & 0 \\
    0 & 0 & 0 & -1 \\
    0 & 0 & 1 & 0
\end{array}\right)
\end{equation}
acts on the moduli as $x_i \to -x_i$, $y_i \to -y_i$ and leaves the RFM invariant. Consequently, it must also be a symmetry of the potential, and derivatives of the Casimir potential with respect to $x_i$ and $y_i$ vanish at $x_i=y_i=0$. The Casimir energy is then given by 
\begin{eqaed}\label{eq:shift}
    \,\mathcal{E}(\mathbf{I})(\alpha) &\propto \sum_{n_1,n_2,m_1,m_2} \frac{1-(-1)^{m_1+m_2}}{(e^{\alpha/\sqrt{2}}(n_1^2+n_2^2)+e^{-\alpha/\sqrt{2}}(m_1^2+m_2^2))^{\frac72}} \,, \\ 
    \,\mathcal{E}(\D[g])(\alpha) &\propto e^{-\alpha (s-2)}\sum_{n_1,n_2} \frac{1}{((n_1+\frac{1}{2})^2+(n_2+\frac{1}{2})^2)^{\frac52}} \,.
\end{eqaed}
The massless spectrum of 7d supergravity fixes the constants of proportionality. Note that the fermions do not contribute to $\mathcal{E}(\D[g])$, so that all phases are trivial in this sum; note also that the sum was projected onto the 2-dimensional subspace invariant under $\D[g]$.
We find a minimum for the total Casimir at $\alpha_{\text{min}} \simeq -0.534$. In this case, the mass matrix adopts a diagonal form, and the squared masses of the moduli are
\begin{align}
    \{m_{x_1}^2,m_{x_2}^2,m_{y_1}^2,m_{y_2}^2,m_{\alpha}^2\}\,L^2 
    = \{-100\,,\,35\,,\,36\,,\,-99\,,\,-44\}\,.
\end{align}
We observe that there are still moduli with negative squared masses below the \ac{BF} bound, $m_{\rm BF}^2L^2 = -1$ \eqref{eq:BF-bound}. 

In the case of the torus, deformations that vary two circles while keeping the overall volume fixed tend to be tachyonic and, as expected, we see this happening here for $x_1$. However, note how this is not the case for $x_2$. This is due to the $1/2$ shift in the twisted term, which makes this example non-generic. This observation will be key in order to construct a perturbatively stable solution.

\subsubsection*{Example 2: $AdS_4 \times T^7/\mathbb{Z}_8$ in M-Theory}

Consider the seven-dimensional RFM with holonomy
\begin{equation}
    \D[g]=\left(\begin{array}{*7{C{1.1em}}}
         0 & -1 & 0 & 0 & 0 & 0 & 0 \\
         1 & 0 & 0 & 0 & 0 & 0 & 0 \\
         0 & 0 & 0 & 0 & 0 & -1 & 0 \\
         0 & 0 & 1 & 0 & 0 & 0 & 0 \\
         0 & 0 & 0 & 1 & 0 & 0 & 0 \\
         0 & 0 & 0 & 0 & 1 & 0 & 0 \\
         0 & 0 & 0 & 0 & 0 & 0 & 1 \\
    \end{array}\right)\,, 
    \quad  \bvec[g]=\left(\begin{array}{c}
         0 \\ 0 \\ 0 \\ 0 \\ 0 \\ 0 \\ \frac{1}{8} \\
    \end{array}\right)
\end{equation}
again, with antiperiodic boundary conditions along the fiber and periodic along the base (see \cite{Aparici:2026wdm} for an extended discussion on this RFM). There are three deformations orthogonal to the volume, which we parametrize as
\begin{equation*}
    \mathbf{g} = R^2 \left(
\begin{array}{ccccccc}
    e^{x-2 y} & 0 & 0 & 0 & 0 & 0 & 0 \\
    0 & e^{x-2 y} & 0 & 0 & 0 & 0 & 0 \\
    0 & 0 & e^{x+y} \cosh (z) & \frac{e^{x+y} \sinh (z)}{\sqrt{2}} & 0 & -\frac{e^{x+y} \sinh (z)}{\sqrt{2}} & 0 \\
    0 & 0 & \frac{e^{x+y} \sinh (z)}{\sqrt{2}} & e^{x+y} \cosh (z) & \frac{e^{x+y} \sinh (z)}{\sqrt{2}} & 0 & 0 \\
    0 & 0 & 0 & \frac{e^{x+y} \sinh (z)}{\sqrt{2}} & e^{x+y} \cosh (z) & \frac{e^{x+y} \sinh (z)}{\sqrt{2}} & 0 \\
    0 & 0 & -\frac{e^{x+y} \sinh (z)}{\sqrt{2}} & 0 & \frac{e^{x+y} \sinh (z)}{\sqrt{2}} & e^{x+y} \cosh (z) & 0 \\
    0 & 0 & 0 & 0 & 0 & 0 & e^{-6 x} \\
\end{array}
\right) \,.
\end{equation*}
Note that when $y=z=0$, the fibre of the RFM becomes isotropic; since derivatives of the Casimir potential with respect to deformations orthogonal to the volume vanish at an isotropic point \cite{Aparici:2025kjj}, we can fix these moduli at this particular point in moduli space. This leaves only $x$ to be stabilised explicitly. The Casimir potential is then extremized at $x_{\rm min}\simeq 0.323$. In this parametrization the mass matrix is diagonal and the moduli have masses given by
\begin{align}
    \{m_x^2,m_y^2,m_z^2\}\,L^2 = \{-131\,,\,-280\,,\,98\} \,.
\end{align}
As with the previous example, there are deformations with squared masses well below the \ac{BF} bound, $m_{\rm BF}^2L^2 = -9/4$ \eqref{eq:BF-bound}. 

What is interesting about this example is that there is no contribution to the Casimir from the group element that generates a $\mathbb{Z}_2$ subgroup within $\mathbb{Z}_8$. This is due to the vanishing of the traces multiplying the lattice sum, and its a general fact for any group element with eigenvalues $(+1)^5(-1)^4$ in the $SO(9)$ little group. Our perturbatively stable solution in Section \ref{sec:stable.ads} is engineered in such a way that all twisted terms in the Casimir potential vanish for this exact reason. This is because, as the following heuristic argument implies, we do not expect tachyon-free vacua to have non-vanishing twisted term contributions to the Casimir.

\subsubsection*{A heuristic no-go argument against twisted terms}\label{sec:heuristic}

The examples above show that finding perturbatively stable AdS is not trivial. Although we do find such a vacuum in Section \ref{sec:stable.ads}, we will spend the remainder of this section providing a heuristic argument as to why a generic RFM inherits tachyons from the parent torus. 

Both examples above contain at least one modulus that is not projected out that parametrizes the relative volume of two blocks in the metric. The existence of such a deformation is always guaranteed in any abelian RFM\footnote{Although this may apply more generally.}, due to the fact that the metric $\mathbf{g}$ must commute with every group element. Since elements in the holonomy group are moduli-independent, it is always possible to find a moduli-independent basis where $\mathbf{g}$ is block-diagonal as we have already seen in Section \ref{sec:inequalityRFM}. Take a group element $\D$ whose invariant subspace is minimal with dimension $k'$, that is, there is no other element with a smaller invariant subspace. Then, there is a basis in which the metric takes the form
\begin{equation}
\mathbf{g} = \begin{pmatrix}
    e^{\frac{2k'}{\sqrt{k k'(k-k')}}x} g_{(k-k')\times(k-k')} & 0 \\
    0 & e^{-\frac{2(k-k')}{\sqrt{kk'(k-k')}}x} G_{k'\times k'} \\
    \end{pmatrix}\,,
\label{metric9090}
\end{equation}
where $x$ is a canonically normalized shape modulus. Within each of the smaller blocks there will be further constraints imposed by the RFM holonomy, but this specific deformation cannot be projected out. This is a consequence of the fact that every element must respect the block diagonal structure of $\mathbf{g}$. 

With this in mind, let us denote by $\mathcal{C}_{\parallel}$ the contribution to the Casimir potential due all group elements whose invariant subspace coincides with that of $\D$ mentioned above. We also introduce the quantity $\mathcal{C}_{\perp} = \mathcal{C} - \mathcal{C}_{\parallel}$, corresponding to every other contribution to the Casimir potential $\mathcal{C}$. Next, we observe that, according to \eqref{eq:Casimir-RFMs},  $\mathcal{C}_{\parallel}$ only depends on the metric \eqref{metric9090} through the bottom-right block $g_{\parallel} = e^{-\frac{2(k-k')}{\sqrt{kk'(k-k')}}x} G_{k'\times k'}$. As a result, we have that
\begin{equation}\label{eq:just.one}
    \mathcal{C}_{\parallel}(x) \propto \frac{1}{|g_{\parallel}|^{\frac{d+k'}{2}}} \propto e^{\frac{(k-k')(d+k')}{\sqrt{kk'(k-k')}}x} \quad \longrightarrow \quad \partial^2_x \mathcal{C}_{\parallel} = \frac{(k-k')(d+k')^2}{kk'} \mathcal{C}_{\parallel}\,.
\end{equation}
Since the smallest eigenvalue of a matrix is smaller than every diagonal entry\footnote{This is also known as the Rayleigh principle \cite{Rayleigh1877}.}, the lowest mass must satisfy 
\begin{equation}
    m^2 L^2 \leq -\frac{k(d-1)^2}{k-d}\frac{\partial_x^2 \mathcal{C}}{\mathcal{C}}.
\end{equation}
Usually, all terms in the Casimir potential satisfy differential equations of the form (\ref{eq:just.one}) as one approaches the asymptotics in moduli space. All solutions we have found are located in the region where these approximate equations are valid. With this in mind, let us assume $\partial_x^2 \mathcal{C}_{\perp}/\mathcal{C} \geq 0$. It follows that
\begin{equation}
    m^2 L^2 \leq -\frac{(k-k')(d+k')^2}{k'}\frac{(d-1)^2}{k-d}\frac{\mathcal{C}_{\parallel}}{\mathcal{C}_{\parallel}+\mathcal{C}_{\perp}}\,.
\end{equation}
Let us define
\begin{equation}
    r=\frac{\mathcal{C}_{\parallel}}{\mathcal{C}_{\parallel}+\mathcal{C}_{\perp}}\,.
\end{equation}
At a critical point, in the absence of cancellations or hierarchies between different terms, we would expect that all terms in the Casimir potential are of the same order of magnitude, so that $r\lesssim 1$. On the other hand, we have that the BF bound will be satisfied if
\begin{equation}
    r \leq r_{\max} \equiv \frac{k'(k-d)}{4(k-k')(d+k')^2}\,.
\end{equation}
In general, this is a small quantity, typically $\mathcal{O}(10^{-3})$ (take e.g. $k=7$, $k'=1$, $d=4$ for which $r_{\max}= 5 \cdot 10^{-3}$). This conflicts with the expectation that $r$ generically attains $O(1)$ values $r\lesssim 1$, and suggests that the deformation under consideration is tachyonic.

This argument hints towards the impossibility of constructing a perturbatively stable vacuum in a compactification where all twisted terms inside $\mathcal{C}_{\parallel}$ are non-zero. Our perturbatively stable solution gets around this precisely by making all these potentially problematic twisted contributions vanish.

\subsection{A perturbatively stable \texorpdfstring{$AdS_4\times T^7/\mathbb{Z}_{2}^{3}$}{AdS4xT7/Z3} with parametric scale separation}
\label{sec:stable.ads}
Here we put into practice everything that we have learned so far. We saw in the previous section that twisted terms generically represent a hurdle when trying to construct tachyon-free vacua. On the other hand, a Casimir energy that is only sourced by the identity gives rise to tachyons, as a deformation that varies the size of two different blocks while keeping the volume constant is expected to be present in any RFM, and these are generically tachyonic\footnote{Note that this is true even in the absence of twisted terms (see e.g. the bare $T^7$ example reviewed in the beginning of \cref{sec:Casimir-compactifications}).}. A way around this would be to give the identity contribution a shift in the lattice sum, as in the twisted term of Example 1 (see the second line of \cref{eq:shift}). Let us see how this can be achieved.

Consider a torus $T^7=\mathbb{R}^7/\Lambda$ where $\Lambda$ is the body-centered cubic lattice, constructed as follows. Let $\{e_i\}$ be a cartesian basis generating a square lattice. A unit cell of the corresponding body centered lattice is generated by the vectors $f_i = e_i$, for $i=1,\ldots,6$ and $f_7 = \sum_{i=1}^7 \frac{1}{2}e_i$. Let us quotient our lattice by the following freely acting group action (expressed in the cartesian basis)
\begin{eqaed}
    \alpha & \, : \, (x_1,x_2,x_3+\frac{1}{2},-x_4,-x_5,-x_6,-x_7)\,,\\
    \beta & \, : \, (x_1+\frac{1}{2},-x_2,-x_3,x_4,x_5,-x_6,-x_7)\,,\\
    \gamma &\, : \, (-x_1,x_2+\frac12,-x_3,x_4+\frac12,-x_5,x_6,-x_7).
\end{eqaed}
This quotient gives rise to an RFM of $\mathbb{Z}_2^3$ holonomy. Interestingly this $\mathbb{Z}_2$ is contained within a $G_2$ subgroup that would preserve a covariantly constant spinor with a periodic spin structure in the covering $T^7$. Indeed, our manifold is closely related to Joyce's $G_2$ toroidal orbifold \cite{Joyce1996_G2}, where shifts have been added in order to avoid orbifold singularities. As for the spin structure, we choose antiperiodic boundary conditions along $f_7$ and periodic along every other direction, i.e. $\vec h = (0,0,0,0,0,0,1/2)$ in the body-centered basis. Furthermore, this quotient projects out all off-diagonal moduli, so that the torus metric written in the cartesian basis becomes
\begin{equation}
    g_{ij} = \text{diag} (e^{2y_1},e^{2y_2},e^{2y_3},e^{2y_4},e^{2y_5},e^{2y_6},e^{2y_7}).
\end{equation}
Deformations that preserve the volume satisfy $\sum_i y_i = 0$.  The holonomy of this RFM has been specifically chosen so that there is no contribution from the twisted terms. Indeed, as explained in Example 2, group elements with eigenvalues $(+1)^5(-1)^4$ on the $SO(9)$ little group have vanishing traces. 
As a consequence, the Casimir energy is comprised exclusively of the identity contribution and takes the schematic form
\begin{equation}
    \mathcal{C} \propto \sum_{\vec \ell\in \Lambda_{\text{bcc}}} \frac{1-e^{2\pi i \vec h \cdot \vec \ell}}{|\vec {\ell}|^{s}} = 2 \sum_{n\in \mathbb{Z}^7} \frac{1}{|\vec n+\frac{1}{2}|^s} \,.
\end{equation}
Here, $\vec \ell$ is expressed in the $f_i$ basis, and its sum runs over all points in the body-centered lattice. On the other hand, $\vec{n}$ runs over the square lattice. We see that, effectively, the chosen spin structure has the effect of giving the identity contribution a shift. Lastly, note that norms are taken with respect to the $g_{ij}$ matrix. 

At the identity point $y_i = 0$, we have that the second derivative with respect to any allowed deformation is negative $\partial^2 \mathcal{C}/\mathcal{C} < 0$. Hence, by eq. \eqref{eq:negative.mass}, it follows that there are no tachyons in the spectrum! In Table \ref{tab:G2RFM}, we display the masses of all relevant deformations, as well as the scaling properties of the solution with the flux number $N$. These scalings are common to any Casimir-flux AdS vacuum and their derivation can be found in Section \ref{sec:Casimir-compactifications}. As a consequence of these powers, the solutions become more and more controlled as $N$ grows, and the 11-dimensional backreaction can be estimated and accounted for as in \cite{ValeixoBento:2025yhz}. In particular, KK modes of every field receive masses of order $1/R$, which is parametrically larger than the AdS scale. 

This vacuum constitutes the first ever scale-separated AdS vacuum with no localized sources in which all moduli are perturbatively stable at one loop. We also do not expect higher loops to destabilize the solution, as all excitations in the spectrum lie well above the BF bound.

\begin{table}[t]
\centering
\small
\begin{tabular}[t]{@{}lc@{}}
\toprule
Mode & $m^2L^2$ \\
\midrule
Shape moduli ($\times6$)   & $+50.248$ \\
Volume modulus ($\times1$) & $+36$ \\
BF bound                   & $-2.25$ \\
\bottomrule
\end{tabular}
\qquad
\begin{tabular}[t]{@{}lc@{}}
\toprule
Scale & $N$-scaling \\
\midrule
Internal radius $R M_{11}$        & $N^{2/3}$ \\
AdS radius $L\,M_{4}$           & $N^{6}$ \\
KK scale $m_{\rm KK}/M_{4}$     & $N^{-3}$ \\
$L^2 m_{\rm KK}^2$                   & $N^{6}$ \\
\bottomrule
\end{tabular}
\caption{Moduli masses and scaling properties of the $AdS_4 \times T^{7}/\mathbb{Z}^3_2$ vacuum.}
\label{tab:G2RFM}
\end{table}

Finally, we wish to comment on the curious trace property that allowed the Casimir energy to vanish in every twisted sector. One might have thought that perhaps this is related to supersymmetry, since our point group would preserve a supercharge with a fully periodic spin structure. However, this is not the case, as we have an antiperiodic circle that breaks all supercharges. Furthermore,  within the twisted terms, we do not observe boson-fermion cancellation; rather, there are separate cancellations within bosons and within fermions, respectively! For the chosen generators, the traces in the $\mathbf{44}$ representation (corresponding to the graviton) and the $\mathbf{84}$ (corresponding to the M-theory three-form) cancel against each other, while the gravitino trace vanishes identically. It would be interesting to understand if there is a deeper reason behind these cancellations, which are also different from the spacetime Atkin-Lehner symmetry introduced in \cite{ValeixoBento:2025yhz}.

\section{Conclusions}
\label{sec:conclus}
In this work, we have found that the simplest approach to constructing lower-dimensional (meta)stable vacua with parametric control and fully stabilized moduli does not work: toroidal compactifications of supergravity supported by Casimir energy and fluxes always lead to tachyonic perturbations violating the \ac{BF} bound. Our results could be extended to more general supergravities, starting from the ten-dimensional case: the presence of the dilaton, Yang-Mills fields and the overall richer structure of these \acp{EFT} could hide some surprises when combined with the ideas explored in this paper. We do not expect such additional fields to qualitatively affect the analysis: if they can be stabilized, the truncated equations of motion reduce to those we have studied. However, if the solution includes kinetic gradients in the internal manifold, the effective potential may change in a significant way.

In any case, as remarked in the Introduction, the techniques we use to obtain this result are more interesting than the results themselves. We have been able to rule out the existence of stable vacua in high-dimensional moduli spaces by showing that the scalar potential satisfies a simple yet non-trivial differential equation in moduli space, and using it to characterize its properties. Thus, our work should be understood in the context of a growing number of papers \cite{Ceresole:1992su, Cecotti:2020uek, Grimm:2020cda, Grimm:2021ikg, Etheredge:2023zjk, Aoufia:2024awo, Basile:2024lcz, Aoufia:2026bau, Aoufia:2026ztl} showing that certain key moduli space quantities are controlled by a local ``moduli space effective field theory'' that can be used to determine the dependence of physical quantities on moduli space coordinates. Our example shows that these quantities can exist even in non-supersymmetric situations. It would be quite transformative if quantities like this could be derived in general setups, as focus could be shifted from direct computation to determining and studying the moduli space differential equations that these objects satisfy.

With that said, the main result of this paper is that we have found the first fully explicit, top-down, perturbatively stable and scale-separated AdS vacuum with no localized sources. We achieved this by studying specific examples of compactifications on several RFMs, which led us to uncover a tachyon-free construction. By compactifying M-theory on a carefully chosen RFM, we managed to project out all tachyonic metric deformations. This happens via non-generic cancellations in the RFM Casimir energy, which deserve further study. Note that the vacuum is not fully stable: there are still non-perturbative decay channels, namely bubble nucleation (studied in detail in \cite{Aparici:2025kjj}), that prevent the existence of a unitary conformal holographic dual. Perturbative stability guarantees a well-defined perturbative EFT, a feature that had not been observed in any other Casimir compactifications until now.

The existence of a perturbatively stable, scale-separated vacuum emphasizes that the current debate about scale-separated vacua is really about the compatibility of this feature with unbroken supersymmetry. Our solution is interesting in a number of ways. One could use it as a starting point for uplifts to interesting solutions of positive vacuum energy. Another avenue involves the exploration of the large $N$ limit of the solution; here, the absence of BF tachyons means that the model might be completed into a non-unitary, non-supersymmetric CFT, as in \cite{Pomoni:2009joh}. In that reference, it was shown that BF tachyons correspond to a loss of conformal invariance, dual to a logarithmic running in the dual CFT. In our case, we might expect something similar, with the caveat that, since the leading instability is non-perturbative, the leading effect breaking conformality would be non-perturbative in $N$. An alternative, proposed in \cite{Antonelli:2018qwz} and developed in \cite{Ghosh:2021lua, Basile:2022zee}, is a unitary but non-conformal holographic dual, undergoing a specific type of renormalization-group flow driven by bubble nucleation. To study this better, it is natural to flux backtrack \cite{Apers:2025pon} this solution, including the Casimir energy, to study the behavior of M2 branes probing the singularity. We hope to come back to some of these interesting questions in the future.

\section*{Acknowledgements}

It is a pleasure to thank David Andriot, Bernardo Fraiman, Hector Parra de Freitas, Lisa Randall, Hassaan Salem, Dimitrios Tsimpis and Cumrun Vafa for discussions. 
The work by MA is supported by the fellowship LCF/BQ/DFR25/12000054 from “La Caixa” Foundation (ID 100010434). The work of IB was supported by the Origins Excellence Cluster and the German-Israel-Project (DIP) on Holography and the Swampland. BB acknowledges the support of the French National Research Agency (ANR) under project ANR-25-CE57-1445-02. BB and MM gratefully acknowledge the support of an Atraccion del Talento Fellowship 2022-T1/TIC-23956 from Comunidad de Madrid, which supported both authors in the early stages of this project, as well as the Spanish State Research Agency (Agencia Estatal de Investigacion) through the grants IFT Centro de Excelencia Severo Ochoa CEX2020-001007-S, RYC2022-037545-I PID2021-123017NB-I00, Europa Excelencia EUR2024-153547, and CEX2025-001574-S, funded by MICIU/AEI/10.13039/501100011033. The research presented in this publication falls under the
research line “Strings and Quantum Gravity”.
We also thank the Aspen Center for Physics in the context of the 2025 Program on Holography, Bootstrap, and Swampland, which is supported by National Science Foundation grant PHY-2210452 and a grant from the Simons Foundation (1161654, Troyer), for providing a stimulating environment during the early stages of this work.

\appendix
\section{Differential equation for the Casimir: alternative derivations}

In this appendix we provide two alternative derivations of the Laplacian eigenvalue equation for the Casimir potential obtained in \secref{sec:PDE-Casimir}, one from a string theory perspective and another based on group theory reasoning.

\subsection{String theory derivation}
\label{ap:PDE-Casimir-strings}

The result of \secref{sec:PDE-Casimir} can also be derived in a stringy way by taking the field-theory limit of a toroidal compactification of string theory (which we take to have vanishing $B$-field). The one-loop vacuum energy is now given by integrating the torus partition function of the worldsheet conformal field theory over the fundamental domain of worldsheet conformal structures, which gives a manifestly \ac{UV}-finite result (even when the vacuum energy is \ac{UV}-sensitive). These conformal structures are parametrized by a complex coordinate $\tau = \tau_1 + i \tau_2$, which replaces the Schwinger proper time $t$, and removes the \ac{UV} region due to modular invariance. Relatedly, the presence of string winding modes along the $k$-torus replaces the heat kernel of \ac{KK} modes with the Narain lattice sum, namely the primary partition function of the worldsheet sigma model,
\begin{eqaed}\label{eq:narain_sum}
    \Gamma_{k,k} = \sum_{n,w \in \mathbb{Z}^k} e^{2\pi i \tau_1 n \cdot w - \pi \tau_2 \left(\frac{1}{R^2}(n \cdot \hat{\mathbf{g}}^{-1} \cdot n) + R^2 \, w \cdot \hat{\mathbf{g}} \cdot w\right)} .
\end{eqaed}
For large $R$ (in units of the string tension) the \ac{KK} spectrum dominates, winding contributions are negligible and the configuration approaches a field-theory limit\footnote{Although it may be constructed, whether a consistent worldsheet background exists is irrelevant for this argument. This approach to derive field-theoretic quantities is in the same spirit as the anomaly analyses of \cite{Schellekens:1986xh}.}. Similarly, the non-compact sector is dominated by the massless modes, which provide a factor of $1/\tau_2^{d/2-1}$. The fundamental domain can be approximated with a strip $\{-\frac{1}{2} \leq \tau_1 \leq \frac{1}{2} \}$ in this limit, and the resulting $\tau_1$-integral becomes trivial\footnote{The integral over $\tau_1$ imposes level matching on worldsheet states, but the contributions we are keeping in the field-theory limit are already level-matched.}. Finally, the modular-invariant measure yields an additional factor of $1/\tau_2^2$, yielding precisely \eqref{eq:cas_pot} where $\tau_2$ plays the role of the Schwinger proper time $t$. In this picture, the differential operator $t^2 \partial_t^2$ in \cref{eq:pde_Q_new,eq:dt_term_pde} can be thought of as the restriction of the Laplacian $\Delta_\tau = \tau_2^2(\partial_{\tau_1}^2+\partial_{\tau_2}^2)$ on the fundamental domain of the worldsheet torus with measure $d^2\tau/\tau_2^2$. Indeed, the Narain lattice sum in \cref{eq:narain_sum} is known to satisfy the partial differential equation \cite{Maloney:2020nni}
\begin{eqaed}
    \left(-\Delta_\tau-\frac{k}{2}\left(1-\frac{k}{2}\right)+\Delta_{\text{Narain}} \right) \left(\tau_2^{\frac{k}{2}}\Gamma_{k,k}\right) = 0 \, ,
\end{eqaed}
which leads to the Laplacian equation for the Casimir energy in the field-theory limit (up to replacing the $SO(k,k)$ Laplacian with the $GL(k)$ one). In fact, the existence of differential equations of this type is generic at the asymptotic boundary of worldsheet conformal manifolds \cite{Aoufia:2024awo, Basile:2024lcz, Aoufia:2026bau}.

\subsection{Group theory derivation}
\label{ap:PDE-Casimir-groups}

Let us begin by recalling some basic facts about toroidal moduli spaces \cite{Obers:1999um}.
The torus $T^k$ is defined by the quotient $\mathbb{R}^k/\mathbb{Z}^k$, with identifications
\begin{equation}
    \vec{x}\sim\vec{x}+\vec{n},\quad\vec{n}\in\mathbb{Z}^k \,.
\label{pres}
\end{equation}
The moduli space of the torus is the space of Riemann-flat metrics on this. Any such metric is described by a $k\times k$ positive-definite, symmetric matrix $\mathbf{g}$, subject to some identifications. Namely, a change of coordinates
\begin{equation}
    \vec{x}\rightarrow \mathbf{M}^{-1}\,\vec{x},\quad \mathbf{M}\in\, SL(k,\mathbb{Z})
\end{equation}
 preserves the identifications \eqref{pres}, and so corresponds to a different parametrization of the same torus. Hence, the corresponding metrics should be identified, leading to the equivalence
\begin{equation}
    \mathbf{g}\,\sim\,\mathbf{M}^T \,\mathbf{g}\,\mathbf{M}.
\label{equiv}
\end{equation}
All of the above is compatible with rescaling the volume of the $T^k$, which henceforth we set to 1. Hence $\text{det}(\mathbf{g})=1$. In this case, a metric on this moduli space should be given by a function quadratic in $d\mathbf{G}$, positive-definite, and invariant under \eqref{equiv}. After restricting to unit volume, this essentially leads to
\begin{equation}
    \frac14\text{Tr}\big((\mathbf{g}^{-1} d\mathbf{g})^2\big) \, ,
\label{erd}
\end{equation}
which is indeed the relevant metric coming from e.g. dimensional reduction of the Einstein-Hilbert action on $T^k$ \cite{Etheredge:2022opl}.

The isometry group of \eqref{erd} includes action of \eqref{equiv} when the matrix $\mathbf{M}$ is taken in $SL(k,\mathbb{R})$. In fact the moduli space of $T^n$ metrics of volume 1, before taking into account the equivalence \eqref{equiv}, and equipped with \eqref{erd}, is a symmetric space of the form $SL(k,\mathbb{R})/SO(k)$. The columns of the $SL(k,\mathbb{R})$ matrix $\mathbf{B}$ can be seen as the basis of the torus lattice in an ambient space equipped with the standard metric. $SO(k)$ in the quotient corresponds to rotations of this basis, and
\begin{equation}
    \mathbf{g}=\mathbf{B}^T\,\mathbf{B} \,.
\end{equation}
At any rate, the metric \eqref{erd} defines a Laplace operator $\Delta$, and we wish to determine the relation of the Casimir energy to this. Since the torus Casimir energy is a lattice sum of the form
\begin{equation}
    \mathcal{C}\sim \sum_{\vec{n}\neq\vec{0}\in\mathbb{Z}^k} \frac{1}{(\vec{n}^T\,\mathbf{g}\,\vec{n})^s} \, ,
\end{equation}
it pays off to understand first the relationship between $\Delta$ and the more elementary function $e_{\vec{n}}\equiv\vec{n}^T\,\mathbf{g}\,\vec{n}$ for fixed $\vec{n}$.

These functions transform linearly under $SL(k,\mathbb{R})$ transformations; an $SL(k,\mathbb{R})$ matrix $\mathbf{N}$ acts on $\mathbf{g}$ as
\begin{equation}
    \mathbf{g}\,\rightarrow \mathbf{N}^T\, \mathbf{g} \,\mathbf{N} \, ,
\end{equation}
so that
\begin{equation}
    e_{\vec{n}}\,\rightarrow\, \vec{n}^T\, \mathbf{N}^T\, \mathbf{g} \,\mathbf{N}\,\vec{n}= e_{\mathbf{N}\vec{n}} \,.
\end{equation}
Thus, the $e_{\vec{n}}$ transform in the symmetric tensor representation of $SL(k,\mathbb{R})$. The Laplace operator $\Delta$ commutes with $SL(k,\mathbb{R})$ transformations, since these are isometries. Therefore, the functions $\Delta e_{\vec{n}}$ also transform in the symmetric tensor representation of $SL(k,\mathbb{R})$; using the explicit expression from the Laplacian, they are also polynomials in the components $\mathbf{g}$. The such function of $\mathbf{g}$ transforming in this representation is precisely the $e_{\vec{n}}$, so it follows that $\Delta$ is a linear transformation on the $e_{\vec{n}}$ that commutes with the $SL(k,\mathbb{R})$ action.
Since the representation is irreducible, Schur's lemma implies that it act by multiplication by a constant:
\begin{equation}
    \Delta (e_{\vec{n}})=\lambda\, e_{\vec{n}}\,.
\end{equation}
and so, the $e_{\vec{n}}$ are eigenfunctions of the Laplacian. In fact, $\Delta$ is equal to the quadratic Casimir operator \cite{Helgason:1984}, as can be constructed from the explicit expression. Thus, the eigenvalue $\lambda$ is simply the value of the quadratic Casimir for the symmetric tensor of $SL(k,\mathbb{R})$, which is (see \cite{Obers:1999um} and substract the radial piece)
\begin{equation}
    \lambda=2\frac{(k-1)(k+2)}{k}
\end{equation}
The Casimir energy is constructed as sums of powers $e_{\vec{n}}^{-s}$ of the elementary functions. The action of the Laplacian,
\begin{equation}
    \Delta(e_{\vec{n}}^{-s})
    = s(s+1)\,\frac{|\nabla e_{\vec{n}}|^2}{\,e_{\vec{n}}^{s+2}}
    - s\,\lambda\,e_{\vec{n}}^{-s} \,.
\label{fr}
\end{equation}
does not manifestly make them eigenfunctions. However the norm of the gradient $e_{\vec{n}}$ turns out to also be proportional to $\vec{e}_{\vec{n}}^2$. Indeed,
\begin{equation}
de_{\vec{n}}= \mathbf{n} \, ,
    \quad \mathbf{n}_{ij}\equiv n_i n_j \,.
\end{equation}
The dual vector to $\mathbf{n}$ is the matrix $4(\mathbf{g}\, \mathbf{n}\,\mathbf{g} -e_n g/k)$, including the volume-corrected term. The norm of the gradient is, therefore,
\begin{equation}
    \frac14\text{Tr}[(4\mathbf{g}^{-1}(\mathbf{g}\, \mathbf{n}\,\mathbf{g} -e_n g/k))^2=4(k-1)e_n^2/k \,.
\end{equation}
Plugging in \eqref{fr}, we discover that also $e_{\vec{n}}^{-s}$ is an eigenfunction of the Laplacian, with eigenvalue
\begin{equation}
    \lambda_s=4\frac{k-1}{k}s(s+1)-\frac{2(k-1)(k+2)}{k}s=\frac{2s(k-1)(2s-k)}{k}.
\end{equation}
 Setting $s=(d+k)/2$ gives $\lambda_s=d(k-1)(d+k)/k$, precisely \eqref{eq:unit-Laplacian-C}.  This also agrees with (2.5a)  of \cite{Obers:1999um}.

\section{Derivation of a Laplace inequality for RFMs}\label{ap:laplace.inequality}

Here we find a Laplace eigenvalue inequality for abelian RFMs, that is however not strong enough to conclude the existence of tachyons. Moving to the Jordan basis of the generators, summing over a rotated lattice $\Lambda_\gamma$, and absorbing the shift vectors $\bvec^\parallel$ into $\vec{n}$ for simplicity of notation, each group element $\gamma\in\Gamma$ contributes to the Casimir energy as
\begin{equation}
    \mathcal{E}(\gamma) = -\hat{\delta}_{\vec{h}}\,\frac{\Gamma(s_\gamma)}{2\pi^{s_\gamma}}\cdot
    \frac{\sqrt{\det\mathbf{g}_\parallel}}{\vert\Gamma\vert}\,\sum_{\vec{n}\in\Lambda_\gamma}\frac{e^{2\pi i\,\theta_{\vec{n}}}}{\left (\sum_a n_i^{(a)}g_{ij}^{(a)}n_j^{(a)}\right )^{s_\gamma}} \,.
\end{equation} 
We have also denoted the phase of each term in the sum by $\theta_{\vec{n}}$, since these moduli independent quantities do not affect the action of the Laplacian on $\mathcal{E}(\gamma)$.
Since the metric $\mathbf{g}_\parallel$ corresponds to the invariant subspace of $\D$ (of dimension $k'$), the sum over metric blocks is appropriately restricted to this subspace.  Acting with the Laplacian \eqref{eq:laplacian}, we find
\begin{align}\label{eq:Laplacian-RFM}
    \hat\Delta\,\mathcal{E}(\gamma) &= 
    \mathcal{E}(\gamma)\Bigg\{k'-\frac{d^2}{k}
        -2s_\gamma\frac{\sum_a h_a\big(n_i^{(a)}g_{ij}^{(a)}n_j^{(a)}\big)}{\left (\sum_ an_i^{(a)}g_{ij}^{(a)}n_j^{(a)}\right )}
        +4s_\gamma(s_\gamma+1)\,\frac{\sum_a f_a\big(n_i^{(a)} g_{ij}^{(a)} n_{j}^{(a)}\big)^2}{\left (\sum_ an_i^{(a)}g_{ij}^{(a)}n_j^{(a)}\right )^{2}}
    \Bigg\} 
\end{align}
where
\begin{align*}
    h_a = \begin{cases}
        3 & a \in A \\
        k_a + 3  & a \notin A 
    \end{cases} \,,
    &&
    f_a = \begin{cases}
        \frac12 & a \in A \\
        1 & a \notin A
    \end{cases}.
\end{align*}
In this form, an obstruction to each $\mathcal{E}(\gamma)$ being an eigenfunction of the Laplacian becomes manifest: for a torus, the sum over blocks in the numerators consists of a single term that cancels against the denominator, but for a general RFM this is no longer the case---some cross terms required for this cancellation are missing in the denominator because in RFMs some metric moduli are projected out. Geometrically, as we shall see more explicitly in the next section, the RFM moduli space is a submanifold of the covering torus moduli space, along which some moduli are fixed and not allowed to vary; their corresponding contributions are absent from the expression for the moduli space Laplacian. 

On the one hand, the Laplacian of $\mathcal{E}(\gamma)$ in \eqref{eq:Laplacian-RFM} can be bounded by noting that\footnote{This follows from\begin{equation}
    \frac{1}{\sum_i^k \frac{1}{a_i} }\leq \frac{\sum_i^k a_i x_i^2}{\left (\sum_i^k x_i \right )^2}\leq \max{a_i}\,, \quad x_i >0 \,.
\end{equation}}
\begin{equation}
    \frac{1}{\ell'}
    \leq\frac{\sum_a f_a \left (n_i^{(a)} g_{ij}^{(a)}n_j^{(a)}\right )^2}{\left (\sum_{a}n_i^{(a)}g_{ij}^{(a)}n_j^{(a)}\right )^{2}}
    \leq 1 \,,
\end{equation}
where $\ell'$ is an effective number of $\D$--invariant blocks that counts rotation blocks twice\footnote{For example, for the identity contribution this would be $\ell'=2\ell_1+\ell_2$.}; and
\begin{equation}
    \min_a \{h_a\}\leq \frac{\sum_a h_a\left (n_i^{(a)} g_{ij}^{(a)}n_j^{(a)}\right )}{{\sum_{a}n_i^{(a)}g_{ij}^{(a)}n_j^{(a)}}}
    \leq \max_a \{h_a\} \,.
\end{equation} 
For any given RFM, all these bounding quantities are known, and one finds the bounds
\begin{align}
    \begin{cases}
        \hat \Delta\,\mathcal{E}(\gamma) \leq \left(k'-\frac{d^2}{k} - 2s_\gamma\max_a\{h_a\} + \frac{4s_\gamma(s_\gamma+1)}{\ell'}\right) \mathcal{E}(\gamma)\,, &\quad\text{if}\quad\mathcal{E}(\gamma)<0 \,, \\
        \hat \Delta\,\mathcal{E}(\gamma) \leq \left(k'-\frac{d^2}{k} - 2s_\gamma\min_a\{h_a\} + 4s_\gamma(s_\gamma+1)\right) \mathcal{E}(\gamma)\,, &\quad\text{if}\quad\mathcal{E}(\gamma)>0 \,.
    \end{cases}
\end{align}
On the other hand, however, the sign of each $\mathcal{E}(\gamma)$ depends on the boundary conditions of the $D$-dimensional fields, encoded in the phases $\theta_{\vec{n}}$; for fully periodic fields, we always have $\mathcal{E}(\gamma)<0$, but for more general boundary conditions, the sign can change.

Moreover, there is one further difficulty to take into account. For RFM compactifications, the Casimir potential \eqref{eq:Casimir-RFMs} is a sum over all $\gamma\in\Gamma$ of definite contributions $\mathcal{E}(\gamma)$ \emph{weighted by the sum of traces} of $\D$ over all massless representations of the $D$-dimensional theory. This means that different terms in $\Vcas$ may have different signs \emph{even if all $\mathcal{E}(\gamma)$ had the same sign}.

Indeed, if all moduli in the internal space metric were allowed to vary, each term in the Casimir potential would become an eigenfunction of the Laplacian. Let $\tilde{\mathcal{E}}(\gamma)$ be a smooth extension of the Casimir potential obtained by allowing all elements of the metric to vary---in the notation introduced above, it is as if there is a single, unrestricted $(a\not\in A)$, $k'$-dimensional block in the metric, i.e. $h_a = k' + 3$ and $f_a = 1$ in \eqref{eq:Laplacian-RFM}. This function then satisfies the differential equation
\begin{align}\label{eq:twisted.generators}
    \hat\Delta\,\tilde{\mathcal{E}}(\gamma) = \lambda_\gamma\,\tilde{\mathcal{E}}(\gamma)\,,
    \quad\text{with}\quad
    \lambda_\gamma = \frac{d(d(k-1)+k(k'-1))}{k}\,,
\end{align}
so that for the full (smoothly extended) Casimir potential we would have
\begin{equation}\label{eq:Laplace-equation-extension}
    \hat\Delta \Tilde{V}_{\rm Cas} = \sum_{\gamma\in \Gamma} \Big(\sum_{\mathbf{r}}{\text{Tr}_\mathbf{r}(\D)}\Big)\hat\Delta\,\tilde{\mathcal{E}}(\gamma) 
    = \sum_{\gamma\in \Gamma} \lambda_\gamma \,\Big(\sum_{\mathbf{r}}{\text{Tr}_\mathbf{r}(\D)}\Big) \tilde{\mathcal{E}}(\gamma) \,. 
\end{equation}
If all terms in this sum were negative, the Laplacian would be bounded by 
\begin{equation}\label{eq:RFM.bound}
    \hat\Delta\Tilde{V}_{\rm Cas} \leq \min \{\lambda_\gamma\} \,\Tilde{V}_{\rm Cas}
\end{equation}
One could then repeat the analysis of Section \ref{sec:no-go-torus}, with $\lambda_c\equiv\min\{\lambda_\gamma\}$, bound the mass of the lightest deformation from above and compare it with the BF bound. As we show in the main text, this happens to be the case for compactifications of maximal supergravity theories.

\section{A geometric perspective on RFMs}\label{ap:geometric.RFM}

We have argued that our no-go for scale separated torus compactifications cannot be directly extended to RFMs, in part because the quotient freezes certain metric moduli of the covering torus. In this section we give a geometric interpretation of this statement. Recall that the moduli space of metrics $\M_\Gamma$ on an RFM of holonomy $\Gamma$ is defined through the condition
\begin{equation}
    \D^T\cdot\mathbf{g}\cdot\D = \mathbf{g} \,.
\end{equation} 
Geometrically, since $\Gamma$ is a freely-acting group of isometries of $T^k$, this condition defines $\M_\Gamma$ as a submanifold of $\M$ corresponding to the fixed-point set of an isometry, 
\begin{equation}
    \M_\Gamma \equiv \{p\in\mathcal{M} : \D^T\cdot\mathbf{g}\cdot\D = \mathbf{g}  \} \,.
\end{equation}
A fixed-point set of an isometry necessarily defines a submanifold that is totally geodesic with respect to $\M$ \cite{Kobayashi1958FixedPO}, which means that any geodesic on $\M_\Gamma$ is also a geodesic on $\M$; in other words, an isometry preserves geodesics and fixes the points of $\M_\Gamma$, forcing geodesics that start tangent to $\M_\Gamma$ to remain in $\M_\Gamma$. 
\begin{figure}
    \centering
    \includegraphics[width=0.7\linewidth]{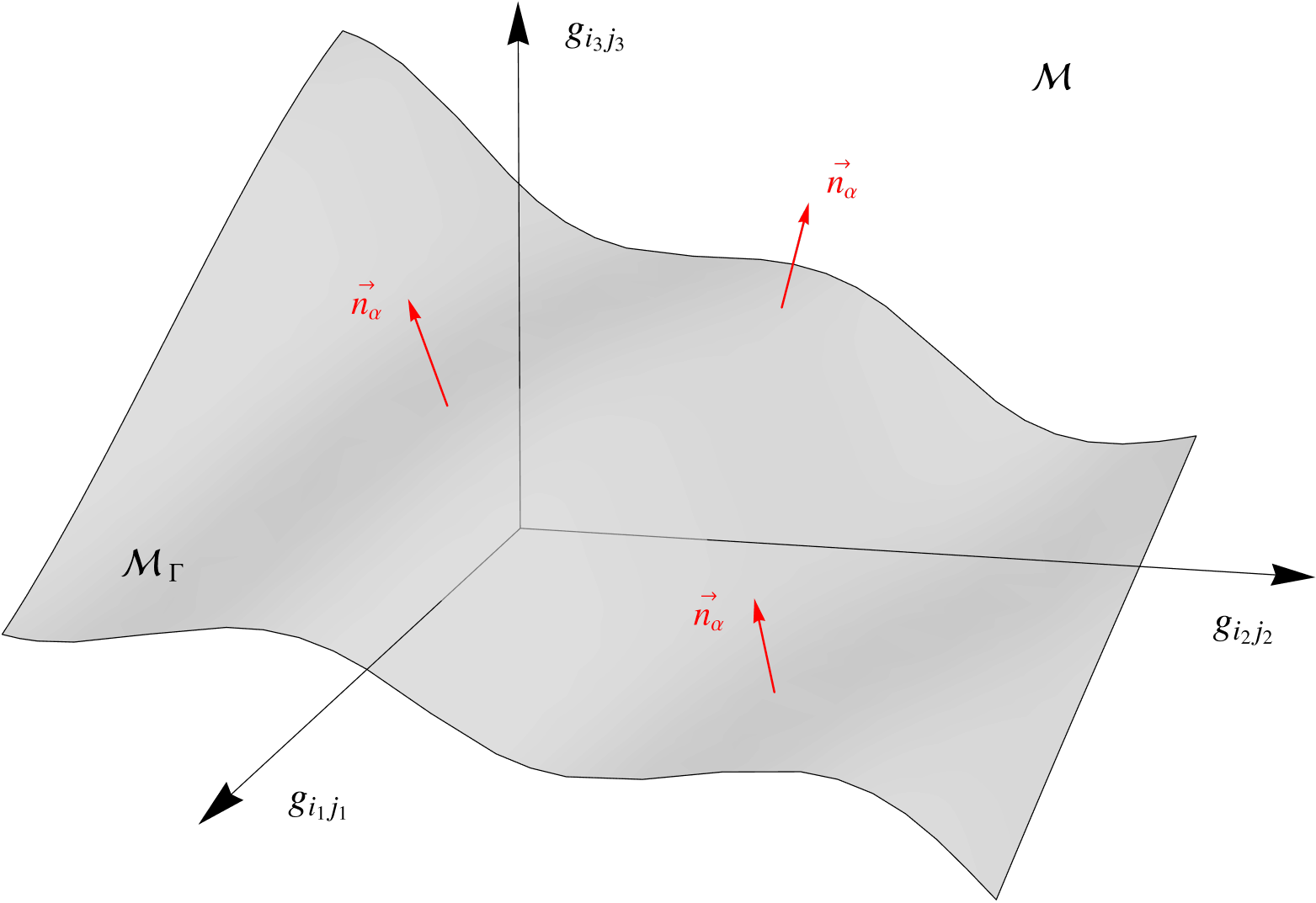}
    \caption{The moduli space of metrics $\M_\Gamma$ on an RFM of holonomy $\Gamma$ corresponds to the fixed-point set of an isometry and is therefore totally geodesic with respect to $\M$. The orthonormal vectors $\vec{n}_{(\alpha)}$ are normal to $\M_\Gamma$ and are associated with directions in moduli space that are frozen by the $\Gamma$--quotient.}
    \label{fig:submanifold}
\end{figure}

We can now relate the action of the Laplacian defined on $\M$ to the action of a Laplacian defined on $\M_\Gamma$. Let $\mathbf{G}$ be the moduli space metric on $\mathcal{M}$, as before, $\mathbf{h}$ be the metric induced on $\M_\Gamma$, and $n_{(\alpha)}^A$ be a set of orthonormal vectors normal to $\M_\Gamma$, where $\alpha = 1,\ldots,q$ for a codimension-$q$ submanifold $\M_\Gamma$ and $A = 1,\ldots,\text{dim}\,\M$. Note that $q$ is the number of metric moduli that are projected out by the $\Gamma$--quotient. Then the Laplacian on $\M_\Gamma$ is related to the Laplacian on $\M$ as 
\begin{equation}
    \Delta_{\M_\Gamma} u_\Gamma = \Delta_\M u \,\big|_{\M_\Gamma} - \sum_\alpha n_{(\alpha)}^A n_{(\alpha)}^B \nabla_A\nabla_B u
    +\sum_\alpha K^\alpha n_{(\alpha)}^A \nabla_A u \,,
\end{equation}
where $u_\Gamma$ is a function on $\M_\Gamma$, $u$ is a smooth extension of $u_\Gamma$ and $K^\alpha$ is the mean curvature defined as 
\begin{equation}
    K^\alpha = \mathbf{h}^{ab} P^A_{\phantom{A}a} P^B_{\phantom{B}b} \nabla_A n^\alpha_B \,,
\end{equation}
in terms of the tangent projector onto $\M_\Gamma$,
\begin{equation}
    P^A_{\phantom{A}B} = \delta^A_{\phantom{A}B} - \sum_\alpha n_{(\alpha)}^A n_{(\alpha)B} \,.
\end{equation} 
Since totally geodesic submanifolds have vanishing extrinsic curvature, we find 
\begin{equation}
    \Delta_{\M_\Gamma} u_\Gamma = \Delta_\M u \,\big|_{\M_\Gamma} - \sum_\alpha n_{(\alpha)}^A n^B_{(\alpha)} \nabla_A\nabla_B u \,,
\end{equation}
for any extension $u$. For our purposes, we want to take $u_\Gamma = \Vcas(\text{RFM})$ as a function of the moduli of an RFM with holonomy $\Gamma$, and $u = \Vcas(T^k)$ as a function of the moduli of the covering $T^k$, i.e. the extension corresponds to allowing all those moduli that are projected out by the action of $\D$ to vary as well---this is what corresponds to eigenfunctions of $\Delta_\M$, cf. \eqref{eq:Laplace-equation-extension}. The difference between these two quantities is then 
\begin{align}
    \Delta_\M \Vcas(T^k) \,\big|_{\M_\Gamma} - \Delta_{\M_\Gamma} \Vcas(\text{RFM}) &= \sum_{\alpha=1}^{\text{codim}(\M_{\Gamma})} n_{(\alpha)}^A n^B_{(\alpha)} \nabla_A\nabla_B \Vcas(T^k) \nonumber \\
    &= \Tr\left(P^\perp\,\nabla_A\nabla_B \Vcas(T^k)\right) \,,
    \label{eq:discrepancy}
\end{align}
with the derivatives taken with respect to the metric $\mathbf{g}$ and where 
\begin{align}
    P^\perp = \sum_{\alpha=1}^{\text{codim}(\M_{\Gamma})} \vec{n}_{(\alpha)}\cdot\vec{n}_{(\alpha)}^T  
\end{align}
is a projector onto the normal subspace spanned by the $T^k$ moduli frozen by the holonomy. \emph{If} the derivatives were evaluated at a saddle point of $\Vcas$ with $\partial_B \Vcas=0$ along all directions, the matrix $\nabla_A\nabla_B \Vcas$ would be precisely the Hessian of $\Vcas$. On the other hand, were we to consider the full potential $V=V_\text{flux} + \Vcas$ and to evaluate the derivatives at a saddle point of $V$, the discrepancy in \eqref{eq:discrepancy} would correspond to the sum over the squared masses of the projected-out moduli, i.e. over the masses these moduli would acquire if they had survived the projection and were stabilised by the potential $V$. However, the saddle point found for an RFM does not typically correspond to a saddle point for the covering torus with the projected-out moduli stabilised at the values fixed by $\Gamma$, and therefore the interpretation is not as clear. More generally, we need the eigenvalues of the matrix 
\begin{equation}
    \nabla_A\nabla_B \Vcas(T^k) = \partial_A\partial_B \Vcas(T^k) - \Gamma^{C}_{AB}\,\partial_C \Vcas(T^k) \,,
\end{equation}
where $\Gamma^{C}_{AB}$ are the connection coefficients with respect to the moduli space metric $\mathbf{G}$. In any case, \eqref{eq:discrepancy} makes it manifest that the failure of $\Vcas$(RFM) to be an eigenfunction of $\Delta_{\M_\Gamma}$ is directly tied to the behaviour of $\Vcas$ along the orthogonal directions frozen by the holonomy. Generalising the proof of Section \ref{sec:no-go-torus} to RFMs would require a bound on this trace that preserves the bounds on $\Delta_\M\Vcas(T^k)$, which would correspond to our previous statement that the holonomy $\Gamma$ does not project out the tachyon guaranteed for $T^k$.

\bibliographystyle{utphys}
\bibliography{bibfilename}

\end{document}